\documentclass[12pt,epsf]{article}
\usepackage[xdvi]{graphicx}
\usepackage{amsmath}
\usepackage{amssymb}
\usepackage{bm}
\usepackage{slashbox}
\usepackage{cite}
\usepackage{comment}
\usepackage{here}

\newcommand{\beq}{\begin{equation}}
\newcommand{\eeq}{\end{equation}}
\newcommand{\bea}{\begin{eqnarray}}
\newcommand{\eea}{\end{eqnarray}}
\newcommand{\non}{\nonumber\\}

\newcommand{\ba}{\begin{array}}
\newcommand{\ea}{\end{array}}
\newcommand{\Slash}[1]{{\ooalign{\hfil/\hfil\crcr$#1$}}}

\newcommand{\cb}{\left|}
\newcommand{\ck}{\right|}
\newcommand{\C}[1]{{\cal{#1}}}
\newcommand{\B}[1]{{\textbf{#1}}}

\newcommand{\T}[1]{{\textrm{#1}}}

\def\lag{{\cal{L}}}
\def\del{\partial}

\def\pe2{p_E^2}

\begin{document}
\setlength{\baselineskip}{0.7cm}
\begin{titlepage}
\begin{flushright}
OCU-PHYS 513  \\
NITEP 39
\end{flushright}
\vspace*{10mm}
\begin{center}{\Large\bf Improving Fermion Mass Hierarchy  \\
\vspace*{2mm}
in Grand Gauge-Higgs Unification \\
\vspace*{2mm}
with Localized Gauge Kinetic Terms}
\end{center}
\vspace*{10mm}
\begin{center}
{\large Nobuhito Maru}$^{a,b}$ and 
{\large Yoshiki Yatagai}$^{a}$, 
\end{center}
\vspace*{0.2cm}
\begin{center}
${}^{a}${\it
Department of Mathematics and Physics, Osaka City University, \\
Osaka 558-8585, Japan}
\\
${}^{b}${\it Nambu Yoichiro Institute of Theoretical and Experimental Physics (NITEP), \\
Osaka City University,
Osaka 558-8585, Japan}
\end{center}
\vspace*{1cm}

\begin{abstract}
Grand gauge-Higgs unification of five dimensional $SU(6)$ gauge theory
 on an orbifold $S^1/Z_2$ with localized gauge kinetic terms is discussed.
The Standard model (SM) fermions on one of the boundaries
 and some massive bulk fermions coupling to the SM fermions on the boundary are introduced,
 so that they respect  an $SU(5)$ symmetry structure.
The SM fermion masses including top quark are reproduced
 by mild tuning the bulk masses and parameters of the localized gauge kinetic terms.
Gauge coupling universality is not guaranteed by the presence of the localized gauge kinetic terms
and it severely constrains the Higgs vacuum expectation value.
Higgs potential analysis shows that the electroweak symmetry breaking occurs
 by introducing additional bulk fermions in simplified representations.
The localized gauge kinetic terms enhance the magnitude of the compactification scale,
 which helps Higgs boson mass large.
Indeed the observed Higgs boson mass 125 GeV is obtained.
\end{abstract}
\end{titlepage}

\section{Introduction}
Gauge-Higgs unification (GHU) \cite{GH} is one of the candidates
 among the physics beyond the Standard Model (SM),
 which solves the hierarchy problem by identifying the SM Higgs field
 with one of the extra spatial component of the higher dimensional gauge field.
In this scenario, the most appealing feature is that physical observables in Higgs sector are calculable
 and predictable regardless of its non-renormalizablity.
For instance, the quantum corrections to Higgs mass and Higgs potential are known to be finite
 at one-loop \cite{1loop} and two-loop \cite{2loop} thanks to the higher dimensional gauge symmetry.
Rich structures of the theory and its phenomenology have been investigated
\cite{Higgsphys, GHUST, diphoton, Maru, GHUflavor, GHUmixing, triple, Yukawa, GHDM}.

The hierarchy problem was originally addressed in grand unified theory (GUT)
 as a problem how the discrepancy between the GUT scale
 and the weak scale are kept and stable under quantum corrections.
Therefore, the extension of GHU to grand unification is a natural direction to explore.
One of the authors discussed a grand gauge-Higgs unification (GGHU) 
 \cite{LM},\footnote{For earlier attempts and related recent works, see \cite{otherGGHU}}
 where the five dimensional $SU(6)$ GGHU was considered
 and the SM fermions were embedded
 in zero modes of $SU(6)$ multiplets in the bulk.
This embedding was very attractive in that it was a minimal matter content
 without massless exotic fermions absent in the SM,
 namely a minimal anomaly-free matter content.
However, a crucial drawback was found.
The down-type Yukawa couplings and the charged lepton Yukawa couplings are not allowed
 since the left-handed $SU(2)_L$ doublets
 and the right-handed $SU(2)_L$ singlets in the down-type sector
 are embedded into different $SU(6)$ multiplets.
As a result, the down-type Yukawa coupling in GHU originated from the gauge coupling cannot be allowed.
This feature seems to be generic in GHU
 as long as the SM fermions are embedded into the bulk fermions.
Fortunately, another approach to generate Yukawa coupling in a context of GHU has been known \cite{CGM, SSS}.
In this approach, the SM fermions are introduced on the boundaries
 (i.e. fixed point in an orbifold compactification).
We also introduce massive bulk fermions, which couple to the SM fermions
 through the mass terms on the boundary.
Integrating out these massive bulk fermions leads to non-local SM fermion masses,
 which are proportional to the bulk to boundary couplings and
 exponentially sensitive to their bulk masses.
Then, the SM fermion mass hierarchy can be obtained by very mild tuning of bulk masses.

Along this line,
 we have improved an $SU(6)$ grand GHU model \cite{LM} in our previous paper \cite{MY},
 where the SM fermion mass hierarchy except for top quark mass is obtained
 by introducing the SM fermions on the boundary as $SU(5)$ multiplets,
 the four types of massive bulk fermions in $SU(6)$ multiplets
 coupling to the SM fermions.
Furthermore, we have shown that the electroweak symmetry breaking and an observed Higgs mass
 can be realized by introducing additional bulk fermions with large dimensional representation.
In GHU, generation of top quark mass is difficult
 since Yukawa coupling is originally gauge coupling and fermion mass is W boson mass as it stands.
The following is well known to overcome this problem
 that if top quark has a mixing with a four rank tensor representation,
 an enhancement of group theoretical factor helps a realization of top quark mass \cite{CCP}.
We have attempted to analyze for the cases of three and four rank tensor representations,
 but an observed top quark mass was not obtained. 

As another known approach \cite{SSS},
 introducing the localized gauge kinetic terms has enhancement effects on fermion masses.
In this paper, we follow this approach.
We consider an $SU(6)$ GGHU model in our previous paper \cite{MY},
 where the SM fermions are localized 4D fields on the boundary
 and the four types of massive bulk fermion.
The localized gauge kinetic terms on the boundaries are added to this model.
Once the localized gauge kinetic terms are introduced,
 the zero mode wave functions of gauge fields are distorted
 and the gauge coupling universality is not guaranteed.
We will find a parameter space where the gauge coupling constant between fermions and a gauge field,
 the cubic and the quartic self-coupling constants are almost universal.
Then, we will show that the fermion mass hierarchy including top quark mass is indeed realized
 by appropriately choosing the bulk mass parameters and the size of the localized gauge kinetic terms.
The correct pattern of electroweak symmetry breaking will be obtained
 by introducing extra bulk fermions as in our previous paper \cite{MY},
 but their representations become greatly simplified.

This paper is organized as follows.
In the next section, we briefly describe the gauge and Higgs sectors of our model.
In section 3, the localized gauge kinetic terms are introduced and
 discuss the mass spectrum of gauge fields including their effects.
In models with the localized gauge kinetic terms,
 the gauge coupling universality is not guaranteed.
We will find a parameter space where the gauge couplings are almost universal.
In section 4, after briefly explaining the generation mechanism of the SM fermion masses,
it is shown that the SM fermion masses including top quark can be reproduced
 by mild tuning of bulk masses and parameters of the localized gauge kinetic terms.
One-loop Higgs potential is calculated and investigated in section 5.
We will show that the observed pattern of the electroweak symmetry breaking
 and Higgs boson mass are realized by introducing some extra bulk fermions.
Final section is devoted to our conclusions.

\section{Gauge and Higgs sector of our model}
In this section, we briefly explain gauge and Higgs sectors of $SU(6)$ GHU model \cite{LM}.
We consider a five dimensional (5D) $SU(6)$ gauge theory
 with an extra space compactified on an orbifold $S^1/Z_2$
 whose radius and coordinate are denoted by $R$ and $y$, respectively.
The orbifold has fixed points at $y=0,\pi R$ and their $Z_2$ parities are given as follows.
 \bea
 P &=& \mbox{diag}(+,+,+,+,+,-) \, \, \mbox{at}~y=0, \non
 P' &=& \mbox{diag}(+,+,-,-,-,-) \, \, \mbox{at}~y=\pi R.
 \eea
We assign the $Z_2$ parity for the gauge field and the scalar field
 as $A_{\mu}(-y)=PA_{\mu}(y)P^\dag$, $A_y(-y)=-PA_y(y)P^\dag$,
 which implies that their fields have the following parities in components,
 \beq
 A_{\mu} = \left(
    \ba{cc|ccc|c}
      (+,+) &(+,+) & (+,-) & (+,-) & (+,-) & (-,-) \\
      (+,+) &(+,+) & (+,-) & (+,-) & (+,-) & (-,-) \\
      \hline
      (+,-) &(+,-) & (+,+) & (+,+) & (+,+) & (-,+) \\
      (+,-) &(+,-) & (+,+) & (+,+) & (+,+) & (-,+) \\
      (+,-) &(+,-) & (+,+) & (+,+) & (+,+) & (-,+) \\
      \hline
      (-,-) &(-,-) & (-,+) & (-,+) & (-,+) & (+,+)
    \ea
  \right),
 \eeq
 \beq
 A_{y} = \left(
    \ba{cc|ccc|c}
      (-,-) &(-,-) & (-,+) & (-,+) & (-,+) & (+,+) \\
      (-,-) &(-,-) & (-,+) & (-,+) & (-,+) & (+,+) \\
      \hline
      (-,+) &(-,+) & (-,-) & (-,-) & (-,-) & (+,-) \\
      (-,+) &(-,+) & (-,-) & (-,-) & (-,-) & (+,-) \\
      (-,+) &(-,+) & (-,-) & (-,-) & (-,-) & (+,-) \\
      \hline
      (+,+) &(+,+) & (+,-) & (+,-) & (+,-) & (-,-)
    \ea
  \right),
 \eeq
where $(+,-)$ means that $Z_2$ parity is even (odd) at $y=0~(y = \pi R)$ boundary, for instance.
We note that only the fields with $(+, +)$ parity has a 4D massless zero mode
 since the wave function takes a form of $\cos(ny/R)$ after the Kaluza-Klein (KK) expansion.
The $Z_2$ parity of $A_\mu$ indicates that $SU(6)$ gauge symmetry is broken
 to $SU(3)_C \times SU(2)_L \times U(1)_Y \times U(1)_X$
 by the combination of the symmetry breaking pattern at each boundary,
 \bea
 &&SU(6)\rightarrow SU(5)\times U(1)_X \, \, \mbox{at}~y=0, \\
 &&SU(6)\rightarrow SU(2)\times SU(4) \, \, \mbox{at}~y=\pi R.
 \eea
The hypercharge $U(1)_Y$ is contained in Georgi-Glashow $SU(5)$ GUT,
 which is an upper-left $5 \times 5$ submatrix of $6 \times 6$ matrix.
Thus, we have a relation of the gauge coupling
\bea
g_3 = g_2 = \sqrt{\frac{5}{3}}g_Y
\eea
at the unification scale, which will not be so far from the compactification scale.
$g_{3,2,Y}$ are the gauge coupling constants for $SU(3)_C, SU(2)_L, U(1)_Y$, respectively.
This coupling relation implies that the weak mixing angle is the same as
 that of Georgi-Glashow $SU(5)$ GUT model, $\sin^2 \theta_W=3/8$~($\theta_W:$weak mixing angle)
 at the unification scale.

The SM $SU(2)_L$ Higgs doublet field is identified
 with a part of an extra component of gauge field $A_y$ as shown below,
 \beq
 A_y=\frac{1}{\sqrt{2}}
  \left(\ba{c|c|c}
  \hspace{30pt}&\hspace{50pt}&H\\ \hline
  &&\\
  &&\\ \hline
  H^{\dag}&&\\
  \ea\right).
 \eeq
We suppose that a vacuum expectation value (VEV) of the Higgs field
 is taken to be in the 28-th generator of $SU(6)$,
 $\langle A_y^a \rangle = \frac{2\alpha}{Rg}\delta^{a\,28}$,
 where $g$ is a 5D $SU(6)$ gauge coupling constant
 and $\alpha$ is a dimensionless constant.
The VEV of Higgs field is given by $\langle H \rangle = \frac{\sqrt{2}\alpha}{Rg}$.
We note that the doublet-triplet splitting problem is solved by the orbifolding
 since the $Z_2$ parity of the colored Higgs field is $(+, -)$
 and it become massive \cite{Kawamura}.

The Higgs couplings to the gauge bosons and the fermons are generated from the gauge interactions,
\begin{equation}
    -\frac{1}{4}\C{F}^a_{MN}\C{F}^{a\,MN}
    \supset
    -\frac{1}{2}\C{F}^a_{\mu y}\C{F}^{a\,\mu y}
    \supset
    -\frac{1}{2}A^a_{\mu}\left(\del_y+f^{adb}A_{y}^d \right)
       \left(\del_y+f^{bec}A_{y}^e\right)A^{\mu\,c},
    \label{gmass}
\end{equation}
\begin{equation}
  \overline{\Psi}i\Slash{D}\Psi
  \supset
  \overline{\Psi}iD_y \Gamma^y \Psi
  =
  -\overline{\Psi}\left(\del_y+A_y\right) \gamma^5 \Psi,
  \label{fmass}
\end{equation}
where
$M$, $N=\{\mu,y\}$ , $\mu=0,1,2,3$ , $y=5$
and subscript $a$,$b$, $c$, $d$, $e$ denote the gauge index for $SU(6)$.
After the Higgs field has the VEV,
eq.~(\ref{gmass}) and eq.~(\ref{fmass}) become the mass terms,
which mass eigenvalues are $m_n(q\alpha)=\frac{n+\nu+q\alpha}{R}$,
where $n$ is KK mode,
$\nu=0$ or $1/2$.
$q$ is an integer charge and
determined by the representation to
which the field with coupling to Higgs field belongs.  
More precisely, the integer charge $q$ is determined by the $SU(2)_L$ representation.
In case where the field with coupling to Higgs field belongs to \B{N+1} representation of $SU(2)_L$,
the integer charge $q$ is equal to $N$.
For instance, in the case of \B{6} representation of $SU(6)$,
since the corresponding branching rule under $SU(6)\rightarrow SU(3)_C\times SU(2)_L$ is
given by $\B{6} \rightarrow (3,1) \oplus (1,2) \oplus (1,1)$,
this representation has four states with $q=0$ and a state with $q=1$.

Some comments on $U(1)_X$ gauge symmetry
 which remains unbroken by orbifolding are given.
We first note that the $U(1)_X$ is in general anomalous 
 since the massless fermions are only the SM fermions
 and their $U(1)_X $ charge assignments are not anomaly-free.
It is easy to cancel the anomaly
 by adding appropriate number of the SM singlet fermions with some $U(1)_X$ charge.
In our model, $U(1)_X$ is supposed to be broken by some mechanism.

\section{Localized gauge kinetic term}
As mentioned in the introduction,
 we introduce localized gauge kinetic terms at $y=0$ and $y=\pi R$ to reproduce a realistic top quark mass.
 Lagrangian for $SU(6)$ gauge field is
  \beq
  \lag_g = \frac{1}{4} \mathcal{F}^{a\,MN}\mathcal{F}^a_{MN}
             - 2 \pi R c_1\delta(y) \frac{1}{4}\mathcal{F}^{b\,\mu\nu}\mathcal{F}^b_{\mu\nu}
                          -2 \pi R c_2\delta(y-\pi R)
                \frac{1}{4}\mathcal{F}^{c\,\mu\nu}\mathcal{F}^c_{\mu\nu},
  \label{lag_g}
  \eeq
  where the first term is the gauge kinetic term in the bulk and $M,N=0,1,2,3,5$.
The second and the third terms are gauge kinetic terms localized at fixed points and $\mu, \nu = 0,1,2,3$.
  $c_{1,2}$ are dimensionless free parameters.
 The subscript $a, b, c$ denote the gauge indices for $SU(6), SU(5)\times U(1), SU(2) \times SU(4)$.
Note that the localized gauge kinetic terms have only to be invariant
 under an unbroken symmetry on each fixed point.

\subsection{Mass spectrum in gauge sector}
Because of the presence of localized gauge kinetic terms,
 the mass spectrum of the SM gauge field becomes very complicated.
In particular, their effects for a periodic sector and an anti-periodic sector are different,
 where the periodic sector means the fields satisfying a condition $A(y+\pi R)=A(y)$
  or those with parity $(P,P')=(+,+), (-,-)$,
  while the anti-periodic sector means the fields satisfying a condition $A(y+\pi R)=-A(y)$
  or those with parity $(+,-), (-,+)$.
This difference originates from the boundary conditions
 for wave functions with a definite charge $q$, $f_n(y; q \alpha)$.
In a basis where 4D gauge kinetic terms are diagonal,
 they are found as $f_n(y+\pi R;q\alpha)=e^{2i\pi q\alpha}f_n(y;q\alpha)$ in periodic sector
 and $f_n(y+\pi R;q\alpha)=e^{2i\pi (q\alpha+1/2)}f_n(y;q\alpha)$ in anti-periodic sector.
Moreover, 
 the wave functions in the same basis satisfy
 \beq
 \left[\del_y^2+m_n^2(q \alpha)
             \left(1+2\pi R c_1\delta(y)+2\pi R c_2 \delta(y-\pi R)\right)
      \right]f_n(y;q \alpha)=0,
 \label{eq:master}
 \eeq
 where $m_n(q \alpha)$ is the KK mass.
By solving eq.~(\ref{eq:master}) with the periodic (anti-periodic) boundary conditions,
 the wave functions and equations determining the KK mass spectrum are obtained \cite{CTW}.
Solving first eq.~(\ref{eq:master}) without boundary terms, we obtain
\beq
 f_n(y;q\alpha) = \mathcal{N}_n(q\alpha+\nu)
    \begin{cases}
        \cos(m_n y) + \beta_n^- \sin(m_n y) \, ,\, y \in [-\pi R,0] \\
        \cos(m_n y) - \beta_n^+ \sin(m_n y) \, ,\, y \in [0,\pi R].  \\
    \end{cases}
    \label{bulksol}
 \eeq
 where $\mathcal{N}_n$ is a normalization factor determined by $\int_{0}^{2\pi R}\cb f_n\ck^2 dy=1$.
$\beta_n^{\pm}$ are integration constants.
Continuity conditions at $y=0, \pi R$ using the above solution $f_n(y; q \alpha)$ lead to
\beq
 \beta_n^{\pm} = e^{\pm i\pi (q\alpha+\nu)} \sec(\pi (q\alpha+\nu))(\pi Rm_n)c_1\mp i \tan(\pi (q\alpha+\nu))\cot(\pi Rm_n)
 \eeq
and eliminating $\beta_n^{\pm}$ in the continuity conditions at $y=0, \pi R$,
 the equations determining the KK mass spectrum
 \beq
 2\left(1-c_1c_2\xi_n^2\right)\sin^2\xi_n+\left(c_1+c_2\right)\xi_n \sin2\xi_n-2 \sin^2(\pi (q\alpha+\nu))=0
 \label{eq:mass}
 \eeq
 is obitaned.
 $\nu$ is $0~(1/2)$ for the periodic (anti-periodic) sector, and $\xi_n=\pi R m_n$.

\begin{figure}[t]
   \centering
   \includegraphics[keepaspectratio, scale=0.7]{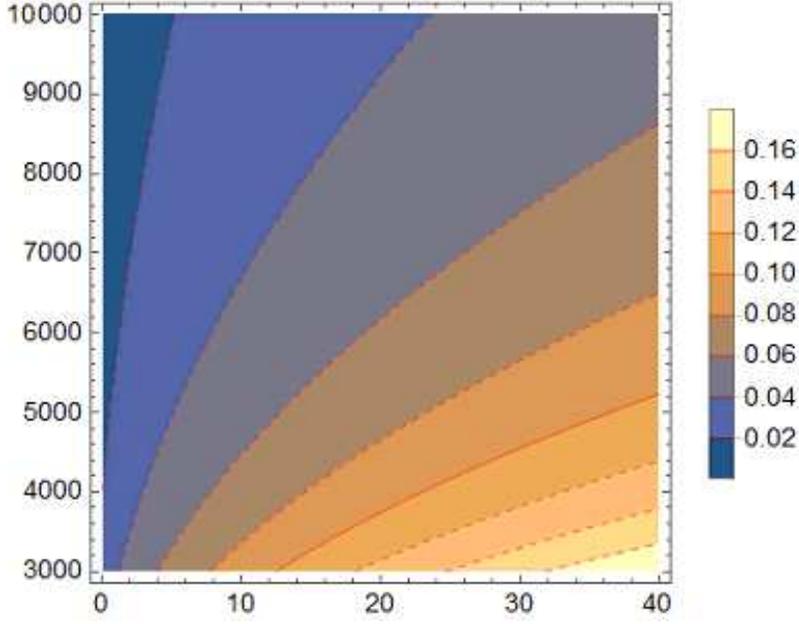}
   \caption{Higgs VEV $\alpha$
   in the range $0 \leq c_1+c_2 \leq 40$ (horizontal axis) and $3000 \mbox{ GeV} \leq 1/R \leq 10000 \mbox{ GeV}$ (vertical axis).}
   \label{figure:Rscale}
  \end{figure}

Since $m_0$ is around weak scale ($\sim 100$ GeV) and $1/R$ is more than 1 TeV,
 it is reasonable to suppose $\xi_0 \ll 1$.
From this observation, we can find an approximate form of $\xi_0$ as
 \beq
 \xi_0 \sim \frac{\sin(\pi (q\alpha+\nu))}{\sqrt{1+c_1+c_2}}.
 \eeq
 For instance,
 the W boson is the gauge boson whose $q$ and $\nu$ are $1$ and $0$, respectively,
 therefore, the W boson mass $m_W$ is given by
 \beq
 m_W = \frac{\sin(\pi\alpha)}{\pi R\sqrt{1+c_1+c_2}}.
 \label{mw}
 \eeq
Moreover, the copmactification scale $1/R$ is determined by this reration when VEV $\alpha$ is obtained (Fig.~\ref{figure:Rscale}) and
 $1/R$ can be large by decreasing $\alpha$.
Since the first KK particle has not observed,
  the copmactification scale should be larger than a few TeV.
Thus, the following condition is needed:
 \beq
 \alpha \leq 0.1.
 \label{VEV}
 \eeq
This condition is very rough estimation.
We would like to emphasize here that $\alpha$ is very small,
which will be important in fitting the fermion mass hierarchy.

\subsection{Gauge coupling universality}
In the SM, the gauge coupling constant between fermions and a gauge boson,
 cubic and quartic self-interaction gauge couplings are universal.
However, in our model,
 the universality of 4D gauge coupling is not maintained
 since the wave functions for massive gauge bosons are distorted from the flat wave functions
 by the localized gauge kinetic terms
 and 4D gauge couplings depend on the integral of the wave functions. 
Therefore, we have to search for a parameter region where the universality is valid.
The gauge coupling between the SM fermions localized at $y=0$ and a 4D gauge boson (KK zero mode: $n=0$)
 is given by
 \beq
 g_{4\,\textrm{gff}}(0;q) = g_5\frac{|f_0(0;q\alpha)|}{\sqrt{Z_0(q\alpha)}}.
 \label{eq:gff}
 \eeq
Similarly, the 4D cubic and quartic self-interaction gauge couplings are given by
 \beq
 g_{4\,\textrm{ggg}}(q_i,q_j,q_k) =
  g_5\int dy [1+2\pi R c_1\delta(y)+2\pi R c_2\delta(y-\pi R)]
  \frac{|f_0(y;q_i \alpha)|}{\sqrt{Z_0(q_i \alpha)}}
  \frac{|f_0(y;q_j \alpha)|}{\sqrt{Z_0(q_j \alpha)}}
  \frac{|f_0(y;q_k \alpha)|}{\sqrt{Z_0(q_k \alpha)}}
 \eeq
 and
 \bea
&& g_{4\,\textrm{gggg}}(q_i,q_j,q_k,q_l) =
  g_5 \left(\int dy [1+2\pi R c_1\delta(y)+2\pi R c_2\delta(y-\pi R)]\right. \nonumber \\
  && \left.\hspace{120pt}\times
  \frac{|f_0(y;q_i \alpha)|}{\sqrt{Z_0(q_i \alpha)}}
  \frac{|f_0(y;q_j \alpha)|}{\sqrt{Z_0(q_j \alpha)}}
  \frac{|f_0(y;q_k \alpha)|}{\sqrt{Z_0(q_k \alpha)}}
  \frac{|f_0(y;q_l \alpha)|}{\sqrt{Z_0(q_l \alpha)}}\right)^{1/2},
 \eea
 where $Z_n(q \alpha)$ is a wave function renormalization factor for the gauge field with a charge $q$
 \beq
 Z_n(q\alpha) = 1 + 2\pi R c_1 \cb f_n(0;q\alpha) \ck^2 + 2\pi R c_2 \cb f_n(\pi R;q\alpha) \ck^2.
 \eeq

\begin{figure}[t]
  \begin{tabular}{cc}
  \begin{minipage}[c]{0.47\hsize}
  \centering
  \includegraphics[keepaspectratio, scale=0.48]{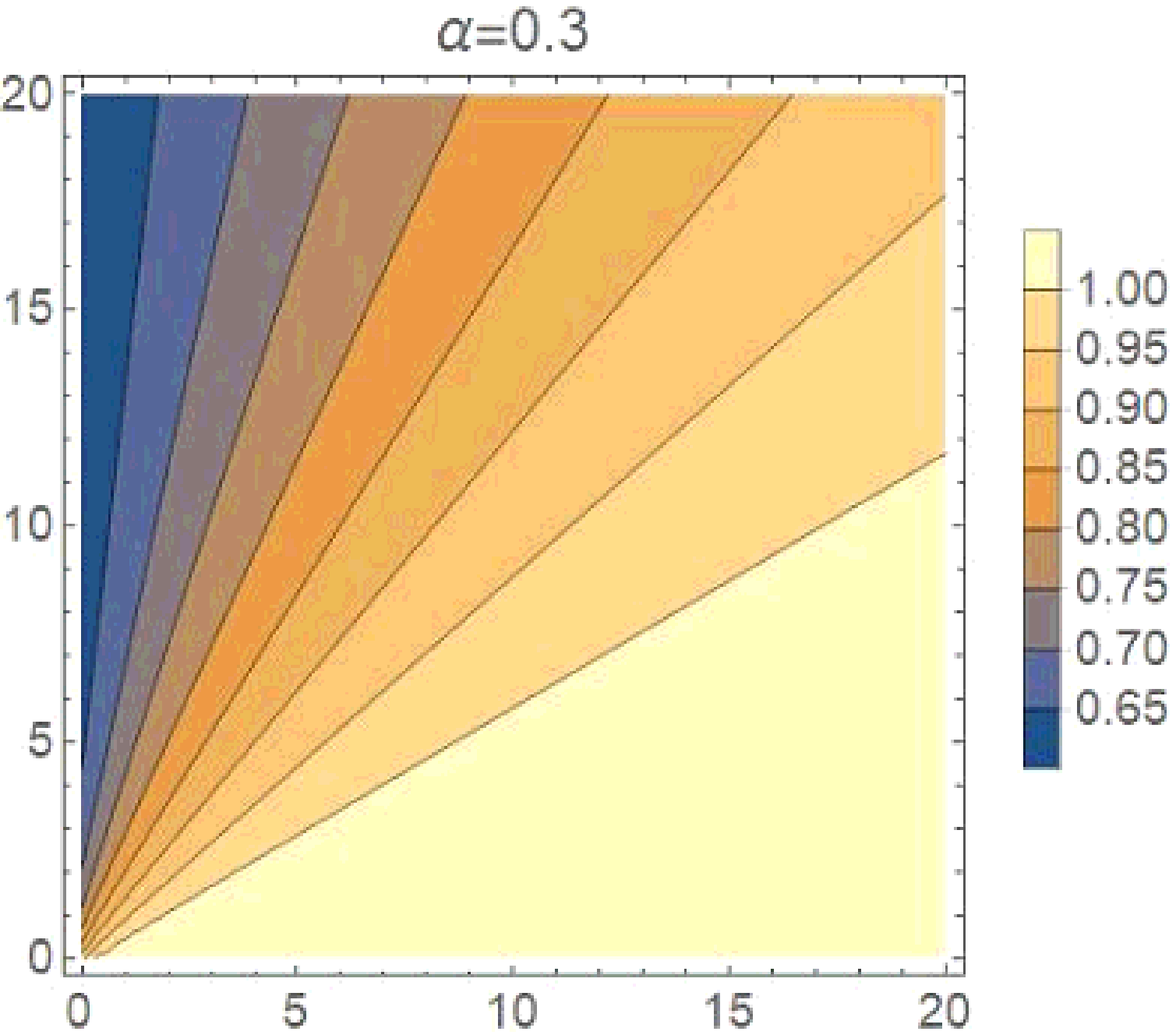}
  \end{minipage}&
  \begin{minipage}[c]{0.47\hsize}
  \centering
  \includegraphics[keepaspectratio, scale=0.48]{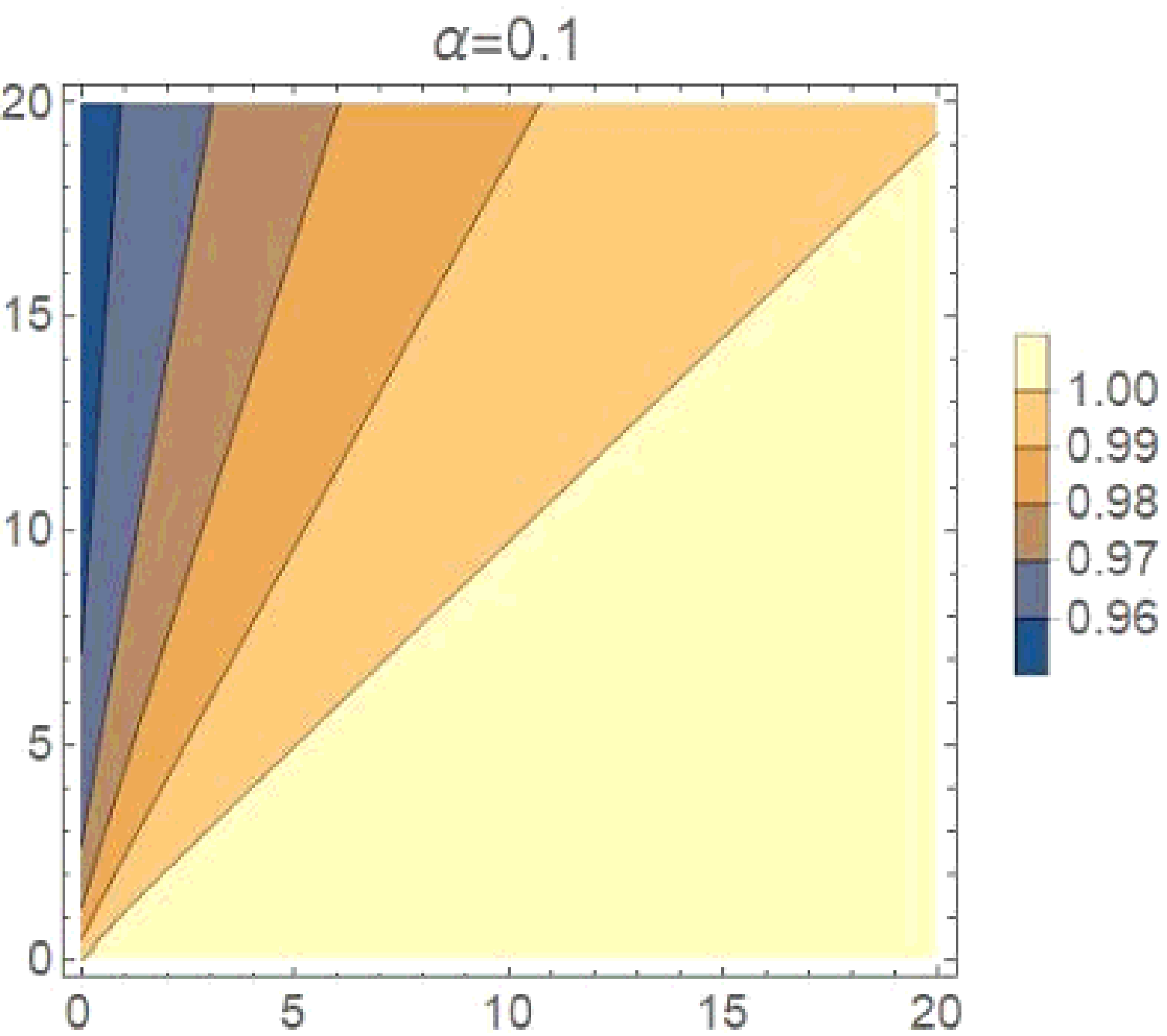}
  \end{minipage}\\
  &\\
  \begin{minipage}[c]{0.47\hsize}
  \centering
  \includegraphics[keepaspectratio, scale=0.47]{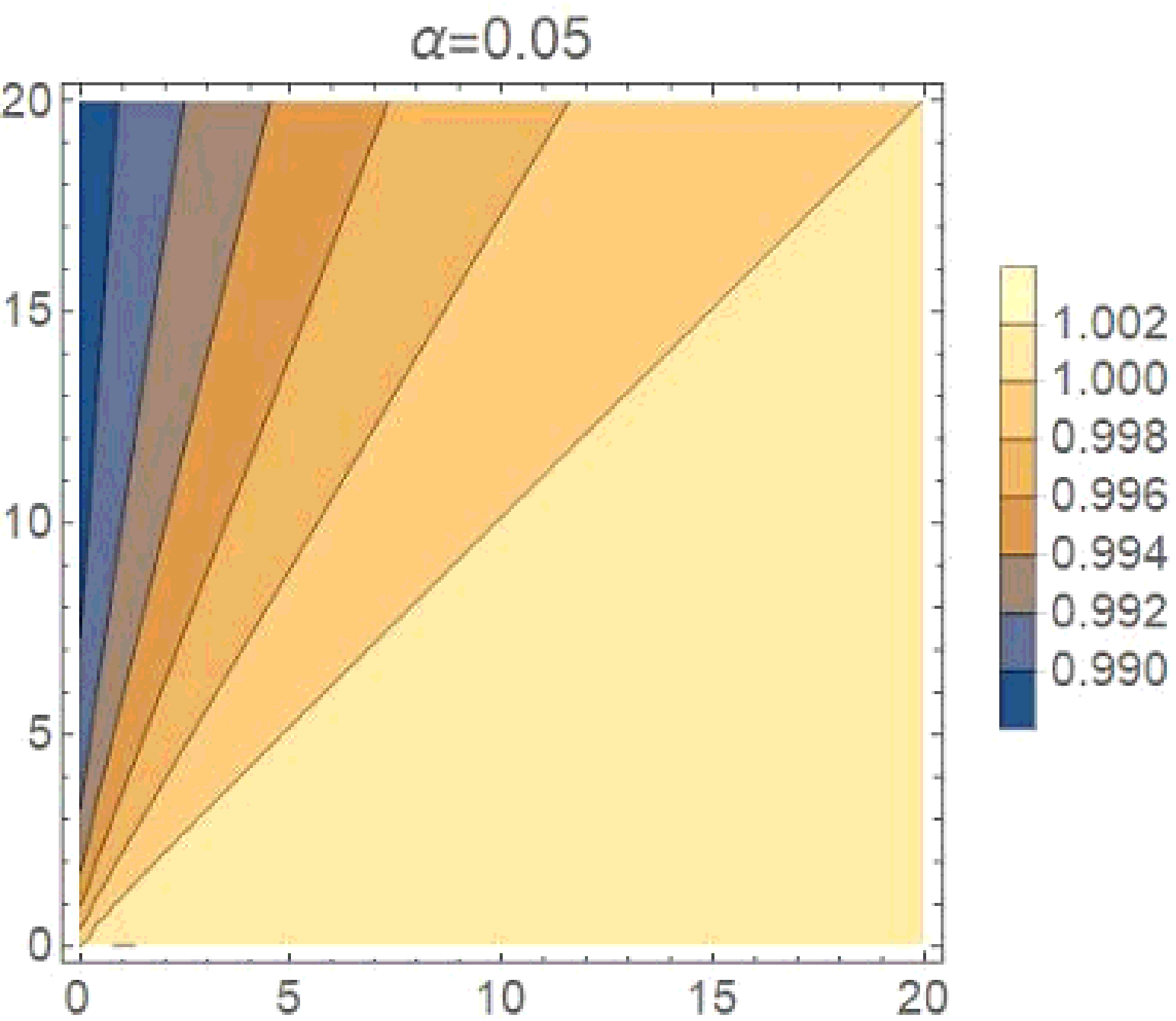}
  \end{minipage}&
  \begin{minipage}[c]{0.47\hsize}
  \centering
  \end{minipage}
  \end{tabular}
  \caption{Ratio between $g_\T{ eff}$ and the effective gauge coupling constant
  between the SM fermions and $q=1$ gauge boson
  in the range $0 \leq c_1 \leq 20$ (horizontal axis) and $0 \leq c_2 \leq 20$ (vertical axis).
  $\alpha$ is taken to be 0.3 (upper left), 0.1 (upper right) and 0.05 (lower left).}
  \label{figure:gaugecouplimg}
 \end{figure}

In the case of $q=0$ corresponding to the photon and the gluon in the SM,
 eq.~(\ref{eq:gff}) is simplified.
According to eq.~(\ref{eq:mass}), we find $m_0(0)=0$,
which implies $f_0(y;0) = \C{N}_n(0) = \frac{1}{\sqrt{2 \pi R}}$
 and $Z_0(0) = 1 + 2\pi R c_1 \cb f_0(0;0) \ck^2 + 2\pi R c_2 \cb f_0(\pi R;0) \ck^2 = 1 + c_1 +c_2$.
Therefore, the gauge coupling universality is valid for $q=0$
 \beq
 g_{4\,\textrm{gff}}(0;0)
 = g_{4\,\textrm{ggg}}(0,0,0)
 = g_{4\,\textrm{gggg}}(0,0,0,0)
 = \frac{g_5}{\sqrt{1+c_1+c_2}\sqrt{2 \pi R}}\equiv g_\T{ eff}.
 \eeq
Then, we have to search for the parameter space
 where the gauge coupling universality is kept for a nonvanishing charge $q$.
Fig.~\ref{figure:gaugecouplimg} shows the ratio between $g_\T{ eff}$
 and the gauge coupling constant between the SM fermions and the W boson ($q=1$). 
The free parameters for the localized gauge kinetic terms are taken in the range $0\leq c_1+c_2 \leq 40$.
In the cases of $\alpha \leq 0.1$ or $1/2 < r = c_1/(c_1+c_2)  < 1$,
 the ratio is almost unity in a good approximation.
As for the cubic and quartic self-interaction gauge coupling constants,
 the ratio is also almost unity in the same parameters.
Since $\alpha$ is restricted to the range $\alpha \leq 0.1$
 which is explained in Section 3.1,
 we do not consider the case $1/2 < r < 1$ hereafter.
After all, the universality of gauge coupling constants can be maintained
 in the range of $\alpha \leq 0.1$. 

\section{Fermion masses}

\subsection{Generation mechanism of the SM fermion masses}
In a paper \cite{LM},
 the SM fermions were embedded in $SU(6)$ multiplets in the bulk,
 which was a minimal matter content
 without massless exotic fermions absent in the SM.
However, the down-type and the charged lepton Yukawa couplings
 are not allowed from $SU(6)$ gauge interactions
 since the left-handed $SU(2)_L$ doublets
 and the right-handed $SU(2)_L$ singlets in the down-type sector
 are embedded into different $SU(6)$ multiplets.
Because of this reason, we take another approach.

The SM quarks and leptons are embedded into $SU(5)$ multiplets localized at $y=0$ boundary,
 which are three sets of decouplet, anti-quintet and singlet $\chi_{10}$, $\chi_{5^{\ast}}$ and $\chi_{1}$.
We also introduce three types of bulk fermions $\Psi$ and $\tilde{\Psi}$  (referred as ``mirror fermions")
 with opposite $Z_2$ parities each other shown in Table~\ref{table:representation}
 and constant mass term such as $M\bar{\Psi} \tilde{\Psi}$
 in the bulk to avoid exotic massless fermions.
Without these mirror fermions and mass terms,
 we necessarily have extra exotic massless fermions with the SM charges after an orbifold compactification.
In this setup, we have no massless chiral fermions from the bulk and its mirror fermions.
The massless fermions are only the SM fermions
 and the gauge anomalies for the SM gauge groups are trivially canceled.
In order to realize the SM fermion masses,
 the boundary localized mass terms between the SM fermions localized at $y=0$ and the bulk fermions are necessary.
To allow such localized mass terms,
 we have to choose appropriate $SU(6)$ representations for bulk fermions carefully.
Note that the mirror fermions have no coupling to the SM fermions.
Table~\ref{table:representation} shows the representations for bulk and mirror fermions introduced in our model
 in addition to the SM fermions, which corresponds to the matter content for one generation.
Totally, three copies of them are present in our model.

\begin{table}[t]
  \centering
   \begin{tabular}{|c|c|} \hline
   bulk fermion $SU(6) \rightarrow SU(5)$
      & mirror fermion \\  \hline
   $20^{(+,P_{20})} = 10 \oplus 10^{\ast}$
      &$ 20^{(-,-P_{20})}$ \\ \hline
   $15^{(+,P_{15})} = 10 \oplus 5$
      &$15^{(-, -P_{15})}$ \\ \hline
   $6^{(-,P_{6})} = 5 \oplus 1$
      &$6^{(+, -P_{6})}$ \\ \hline
  \end{tabular}
  \vspace{-5pt}
 \caption{Representation of bulk fermions and
 the corresponding mirror fermions.
 $P_i$ are parity of bulk fermion for \B{i} representation in $SU(6)$
 ($P_i=\pm 1$).
 $R$ in $R^{(+,+)}$ means an $SU(6)$ representation of the bulk fermion.
 $r_i$ in $r_1 \oplus r_2$ are $SU(5)$ representations.}
 \label{table:representation}
 \vspace{20pt}
  \centering
   \begin{tabular}{|c|c|} \hline
   bulk fermion $SU(5) \rightarrow SU(3)_C \times SU(2)_L \times U(1)_Y$
         & SM fermion coupling to bulk \\  \hline
   \scalebox{0.9}{
   $10 =
       Q_{20}(3,2)_\T{{1/6}}^{(+,P_{20})}
        \oplus
         U^{\ast}_{20}(3^{\ast},1)_\T{{-2/3}}^{(+,-P_{20})}
        \oplus
           E^{\ast}_{20}(1,1)_\T{{1}}^{(+,-P_{20})}$}

      &\scalebox{0.9}{
      $
         q_L(3,2)_\T{{1/6}},
         u_R^c(3^{\ast},1)_\T{{-2/3}},
         e_R^c(1,1)_\T{{1}}$}
          \\ \hline
   \scalebox{0.9}{
   $10^{\ast}  =
        Q_{20}^{\ast}(3^{\ast},2)_\T{{-1/6}}^{(-,-P_{20})}
         \oplus
        U_{20}(3,1)_\T{{2/3}}^{(-,P_{20})}
         \oplus
        E_{20}(1,1)_\T{{-1}}^{(-,P_{20})}$}
     &\scalebox{0.9}{$
        q_L^c(3^{\ast},2)_\T{{-1/6}},
        u_R(3,1)_\T{{2/3}},
        e_R(1,1)_\T{{-1}}$} \\ \hline
  \end{tabular}
  \vspace{-5pt}
 \caption{\B{20} bulk fermion and SM fermions per a generation.
 $P_{20}$ is parity of bulk fermion for \B{20}
 ($P_{20}=\pm 1$).
 $R$ in $R^{(+,+)}$ means an $SU(6)$ representation of the bulk fermion.
 $r_{1,2}$ in $(r_1, r_2)_{a}$ are $SU(3)$, $SU(2)$ representations in the SM, respectively.
 $a$ is U(1)$_Y$ charges.}
 \label{table:representation20}
 \vspace{20pt}
  \centering
   \begin{tabular}{|c|c|} \hline
   bulk fermion $SU(5) \rightarrow SU(3)_C \times SU(2)_L \times U(1)_Y$
      & SM fermion coupling to bulk \\  \hline
   \scalebox{0.9}{
   $10 = Q_{15}(3,2)_\T{{1/6}}^{(+,-P_{15})} \oplus
         U^{\ast}_{15}(3^{\ast},1)_\T{{-2/3}}^{(+,P_{15})} \oplus
         E^{\ast}_{15}(1,1)_\T{{1}}^{(+,P_{15})}$}
      &\scalebox{0.9}{
      $ q_L(3,2)_\T{{1/6}},
         u_R^c(3^{\ast},1)_\T{{-2/3}},
         e_R^c(1,1)_\T{{1}}$} \\ \hline
  $5 = D_{15}(3,1)_\T{{-1/3}}^{(-,P_{15})} \oplus
         L^{\ast}_{15}(1,2)_\T{{1/2}}^{(-,-P_{15})}$
      &$ d_R(3,1)_\T{{-1/3}},
         l_L^c(1,2)_\T{{1/2}}$ \\ \hline
  \end{tabular}
  \vspace{-5pt}
 \caption{\B{15} bulk fermion and SM fermions per a generation.
 $P_{15}$ is parity of bulk fermion for \B{15}
 ($P_{15}=\pm 1$).
 $r_{1,2}$ in $(r_1, r_2)_{a}$ are $SU(3)$, $SU(2)$ representations in the SM, respectively.
 $a$ is U(1)$_Y$ charges.}
 \label{table:representation15}
 \vspace{20pt}
  \centering
   \begin{tabular}{|c|c|} \hline
   bulk fermion $SU(5) \rightarrow SU(3)_C \times SU(2)_L \times U(1)_Y$
      & SM fermion coupling to bulk \\  \hline
   $5 = D_{6}(3,1)_\T{{-1/3}}^{(-,-P_{6})} \oplus
         L^{\ast}_{6}(1,2)_\T{{1/2}}^{(-,P_{6})}$
      &$ d_R(3,1)_\T{{-1/3}},
         l_L^c(1,2)_\T{{1/2}}$ \\ \hline
   $1 = N_{6}^{\ast}(1,1)_\T{{0}}^{(+,-P_{6})} $
      &$ \nu_R^c(1,1)_\T{{0}}$ \\ \hline
  \end{tabular}
  \vspace{-5pt}
 \caption{\B{6} bulk fermion and SM fermions per a generation.
 $P_{6}$ is parity of bulk fermion for \B{6}
 ($P_{6}=\pm 1$).
 $r_{1,2}$ in $(r_1, r_2)_{a}$ are $SU(3)$, $SU(2)$ representations in the SM, respectively.
 $a$ is U(1)$_Y$ charges.}
 \label{table:representation15}
 \end{table}

Lagrangian for the fermions is given by
\bea
\lag_{\T{ matter}}&=&\sum_{a=20, 15, 6} \left[\overline{\Psi}_ai\Gamma^MD_M\Psi_a
                  +\overline{\tilde{\Psi}}_ai\Gamma^MD_M\tilde{\Psi}_a
                      + \left( \frac{ \lambda_a}{\pi R}\overline{\Psi}_a \tilde{\Psi}_a + \T{ h.c.} \right) \right]\non
&&+\delta(y)\left[\overline{\chi}_{10}i\Gamma^{\mu}D_{\mu}\chi_{10}
                     +\overline{\chi}_{5^{\ast}}i\Gamma^{\mu}D_{\mu}\chi_{5^{\ast}}
                     +\overline{\chi}_1i\Gamma^{\mu}D_{\mu}\chi_1\right.\non
&&
     +\sqrt{\frac{2}{\pi R}}\left\{
          \epsilon_{20} \left(\overline{\chi}_{10}\Psi_{10\subset 20} + \overline{\chi}_{10}^{c}\Psi_{10^{\ast} \subset 20} \right)
        +\epsilon_{15} \left(\overline{\chi}_{10}\Psi_{10\subset 15} + \overline{\chi}_{5^{\ast}}^{c}\Psi_{5\subset 15} \right) \right.\non
&&\left.\left.
     +\epsilon_6 \left(\overline{\chi}_{5^{\ast}}\Psi_{5\subset 6} + \overline{\chi}_{1}\Psi_{1\subset 6} \right)
        +\mbox{h.c.}
      \right\} \right],
\eea
where
the five-dimensional gamma matrices $\Gamma^M$ is given
by $(\Gamma^{\mu},\Gamma^y)=(\gamma^{\mu},i\gamma^5)$.

The first line is Lagrangian for the bulk and mirror fermions,
 and the remaining terms are Lagrangian localized on $y=0$ boundary.
Note that the subscript $``a"$ denotes the $SU(6)$ representations of the bulk and mirror fermions.
The bulk masses between the bulk and the mirror fermions are normalized by $\pi R$
 and expressed by the dimensionless parameter $\lambda_a$.
The last two lines are mixing mass terms between the bulk fermions and the SM fermions.
$\Psi_{M\subset N}$ is a bulk fermion for $M$ in $SU(5)$ representation
 and $N$ means $SU(6)$ representation.
$\epsilon_i$ are the strength of the mixing term between the bulk fermion and the SM fermion.
Note that all of the boundary terms respect $SU(5)$ GHU symmetry structure.
Integrating out $y$-direction after KK expansion of bulk fermions
 leads to the following 4D effective Lagrangian.
 \bea
 \mathcal{L}_4 &\supset& \sum_{n = - \infty}^{\infty}
            \Big[ \overline{\Psi}^{(n)} (i \Slash{\partial} - m_n(q\alpha)) \Psi^{(n)}
            + \overline{\tilde{\Psi}}^{(n)} (i \Slash{\partial} + m_n(q\alpha)) \tilde{\Psi}^{(n)} \non
 &&
 + \left.\left( \frac{\lambda}{\pi R} \overline{\Psi}^{(n)} \tilde{\Psi}^{(n)}
            +  \epsilon_i \overline{\psi_{\T{ SM}}} \frac{\kappa_L P_L + \kappa_R P_R}{\pi R} \Psi^{(n)}
            + \mbox{h.c.} \right) \right],
\label{4dL}
 \eea
 where $\Psi^{(n)}(\tilde{\Psi}^{(n)})$ represents a $n$-th KK mode of bulk (mirror) fermion,
 and $\psi_{\T{ SM}}$ is a SM fermion.
$P_{L,R}$ are chiral projection operators and $\kappa_{L,R}$ are some constants,
in this model $\kappa_{L,R}=0$ or $1$.
$m_n(q\alpha) = \frac{n + \nu + q\alpha}{R}$ denotes the sum of the ordinary KK mass
 and the electroweak symmetry breaking mass proportional to the Higgs VEV $\alpha$
 where $\nu=0$ or $1/2$.
The charge $q$ is determined by the representation which the fermion 
 belongs to.
The mass spectrum of bulk and mirror fermions is totally given
 by $m_n^2 = \left(\frac{\lambda}{\pi R}\right)^2 + m_n(q\alpha)^2$.
Note that the Lagrangian (\ref{4dL}) is illustrated for a particular bulk and mirror fermion as an example.

A comment on the bulk mass spectrum
 $m_n^2 = \left(\frac{\lambda}{\pi R}\right)^2 + m_n(q\alpha)^2$ is given.
This spectrum is not exactly correct in the case
 that the mixings between the bulk and the boundary fermions are large.
Following the argument in \cite{SSS},
 we also assume in this paper that the physical mass induced for the boundary fields
 is much smaller than the masses of the bulk fields.
In this case, the effects of the mixing on the spectrum for the bulk fields can be negligible
 and the spectrum $m_n^2 = \left(\frac{\lambda}{\pi R}\right)^2 + m_n(q\alpha)^2$ is a good approximation.

In order to derive the SM fermion masses,
 we need the quadratic terms in the effective Lagrangian for the SM fermion.
 \bea
 {\cal L}_{\T{ SM}} \supset \overline{\psi}_{\T{ SM}} \sum_i K_i \psi_{\T{ SM}}
 \eea
with
 \beq
 \label{eq:K}
 K_i \equiv \Slash{p} \left( 1 + \epsilon_i^2 \frac{\kappa_L^2 P_L + \kappa_R^2 P_R}{\sqrt{x^2 + \lambda_i^2}} \right)
 \mbox{Re} f_0^{(\pm P_i)}(\sqrt{x^2 + \lambda^2_i}, q \alpha)
    +\frac{\epsilon_i^2\kappa_L \kappa_R}{\pi R} \mbox{Im} f_0^{(\pm P_i)}(\sqrt{x^2 + \lambda^2_i}, q \alpha)
 \eeq
where $x \equiv \pi R p$ and
 \beq
 f_0^{(+)}(\sqrt{x^2 + \lambda^2}, q \alpha)
 \equiv \sum^{\infty}_{n = - \infty} \frac{1}{\sqrt{x^2 + \lambda^2} + i \pi (n + q \alpha)}
 = \mbox{coth}(\sqrt{x^2 + \lambda^2} + i \pi q\alpha),
 \eeq
 \beq
 f_0^{(-)}(\sqrt{x^2 + \lambda^2}, q \alpha)
 \equiv \sum^{\infty}_{n = - \infty} \frac{1}{\sqrt{x^2 + \lambda^2} + i \pi (n+ 1/2 + q \alpha)}
 = \mbox{tanh}(\sqrt{x^2 + \lambda^2} + i \pi q\alpha).
 \eeq
In deriving ${\cal L}_{\T{ SM}}$,
 we simply took the large bulk mass limit $\frac{\lambda^2}{(\pi R)^2} \gg p^2$
 so that the mixings of the SM fermions with non-zero KK modes become negligibly small.

Integrating out all massive bulk fermions and normalizing the kinetic term to be canonical,
 we obtain the physical mass for the SM fermions.
 \beq
 m^a_{\T{ phys}} = \left. \frac{m^a}{\sqrt{Z_L^a Z_R^a}} \right|_{x \ll \lambda}
 \simeq \sqrt{1+c_1+c_2} m_W e^{- \lambda}~(a=u, d, e, \nu)
 \label{physmass}
 \eeq
where the bare mass and the wave function renormalization factors are
\bea
m^a &=& \sum_i \frac{\epsilon_i^2\kappa_L^i\kappa_R^i}{\pi R} \mbox{Im} f_0^{(\pm P_i)}(\sqrt{x^2 + \lambda^2_i}, q_i \alpha) ,
\label{m}\\
Z^a_{L, R} &=& 1+ \sum_i \frac{\epsilon_i^2 \kappa^i_{L,R}}{\sqrt{x^2 + \lambda_i^2}}
 \mbox{Re} f_0^{(\pm P_i)}(\sqrt{x^2 + \lambda_i^2}, q_i \alpha).
\label{Z}
\eea
The summation in $Z^a_{L, R}$ is taken
 for all the bulk fields contributing to mass $m^a$
 and its precise expressions are explicitly shown in the next subsection.

In  this model, these are explicitly given
\begin{equation}
  \begin{aligned}
    m^{u} &
     = \frac{\epsilon_{\T{20}}^2}{\pi R}
     \left[\mbox{Im}f_0^{(P_{20})}(\sqrt{x^2+\lambda_{20}^2},\alpha)
      + \mbox{Im}f_0^{(-P_{20})}(\sqrt{x^2+\lambda_{20}^2},\alpha)
      \right],
      \\
    Z^{u}_L &
     = 1
       +\frac{\epsilon_{\T{20}}^2}{\sqrt{x^2 + \lambda_{20}^2}}
       \left[
         \mbox{Re}f_0^{(P_{20})}(\sqrt{x^2+\lambda_{20}^2},\alpha)
          +\mbox{Re}f_0^{(-P_{20})}(\sqrt{x^2+\lambda_{20}^2},\alpha)
          \right]\\
     &\hspace{21pt}
       +\frac{\epsilon_{\T{15}}^2}{\sqrt{x^2 + \lambda_{15}^2}}
          \mbox{Re}f_0^{(-P_{15})}(\sqrt{x^2+\lambda_{15}^2},0),\\
    Z^{u}_R &
      = 1
        +\frac{\epsilon_{\T{20}}^2}{\sqrt{x^2 + \lambda_{20}^2}}
        \left[
         \mbox{Re}f_0^{(P_{20})}(\sqrt{x^2+\lambda_{20}^2},\alpha)
         +\mbox{Re}f_0^{(-P_{20})}(\sqrt{x^2+\lambda_{20}^2},\alpha)
         \right]\\
      &\hspace{21pt}
        +\frac{\epsilon_{\T{15}}^2}{\sqrt{x^2 + \lambda_{15}^2}}
           \mbox{Re}f_0^{(P_{15})}(\sqrt{x^2+\lambda_{15}^2},0),\\
   &\\
  \end{aligned}
\end{equation}
\begin{equation}
  \begin{aligned}
    m^{d} &
     = \frac{\epsilon_{\T{15}}^2}{\pi R}\mbox{Im}f_0^{(-P_{15})}(\sqrt{x^2+\lambda_{15}^2},\alpha),
      \\
    Z^{d}_L &
     = 1
       +\frac{\epsilon_{\T{20}}^2}{\sqrt{x^2 + \lambda_{20}^2}}
       \left[
          \mbox{Re}f_0^{(P_{20})}(\sqrt{x^2+\lambda_{20}^2},0)
         +\mbox{Re}f_0^{(-P_{20})}(\sqrt{x^2+\lambda_{20}^2},0)
         \right]\\
       &\hspace{21pt}
       +\frac{\epsilon_{\T{15}}^2}{\sqrt{x^2 + \lambda_{15}^2}}
          \mbox{Re}f_0^{(-P_{15})}(\sqrt{x^2+\lambda_{15}^2},\alpha),\\
    Z^{d}_R &
      = 1
        +\frac{\epsilon_{\T{15}}^2}{\sqrt{x^2 + \lambda_{15}^2}}
           \mbox{Re}f_0^{(-P_{15})}(\sqrt{x^2+\lambda_{15}^2},\alpha)
        +\frac{\epsilon_{\T{6}}^2}{\sqrt{x^2 + \lambda_6^2}}
           \mbox{Re}f_0^{(P_{6})}(\sqrt{x^2+\lambda_6^2},0),\\
   &\\
  \end{aligned}
\end{equation}
\begin{equation}
  \begin{aligned}
    m^{e} &
     = \frac{\epsilon_{\T{15}}^2}{\pi R}\mbox{Im}f_0^{(P_{15})}(\sqrt{x^2+\lambda_{15}^2},\alpha),
      \\
    Z^{e}_L &
     = 1
       +\frac{\epsilon_{\T{15}}^2}{\sqrt{x^2 + \lambda_{15}^2}}
        \mbox{Re}f_0^{(P_{15})}(\sqrt{x^2+\lambda_{15}^2},\alpha)
      +\frac{\epsilon_{\T{6}}^2}{\sqrt{x^2 + \lambda_6^2}}
         \mbox{Re}f_0^{(-P_{6})}(\sqrt{x^2+\lambda_6^2},0),\\
    Z^{e}_R &
      = 1
        +\frac{\epsilon_{\T{20}}^2}{\sqrt{x^2 + \lambda_{20}^2}}
        \left[
         \mbox{Re}f_0^{(P_{20})}(\sqrt{x^2+\lambda_{20}^2},0)
         +\mbox{Re}f_0^{(-P_{20})}(\sqrt{x^2+\lambda_{20}^2},0)
         \right]\\
       &\hspace{21pt}
        +\frac{\epsilon_{\T{15}}^2}{\sqrt{x^2 + \lambda_{15}^2}}
         \mbox{Re}f_0^{(P_{15})}(\sqrt{x^2+\lambda_{15}^2},\alpha),\\
  \end{aligned}
\end{equation}
\begin{equation}
  \begin{aligned}
    m^{\nu} &
     = \frac{\epsilon_{\T{6}}^2}{\pi R}\mbox{Im}f_0^{(-P_6)}(\sqrt{x^2+\lambda_6^2},\alpha),
      \\
    Z^{\nu}_L &
     = 1
       +\frac{\epsilon_{\T{15}}^2}{\sqrt{x^2 + \lambda_{15}^2}}
        \mbox{Re}f_0^{(P_{15})}(\sqrt{x^2+\lambda_{15}^2},0)
       +\frac{\epsilon_{\T{6}}^2}{\sqrt{x^2 + \lambda_6^2}}
          \mbox{Re}f_0^{(-P_{6})}(\sqrt{x^2+\lambda_6^2},\alpha),\\
    Z^{\nu}_R &
      = 1
        +\frac{\epsilon_{\T{6}}^2}{\sqrt{x^2 + \lambda_6^2}}
           \mbox{Re}f_0^{(-P_6)}(\sqrt{x^2+\lambda_6^2},\alpha).
  \end{aligned}
\end{equation}
Focusing on parity for \B{20},
 since $P_{20}=\pm 1$ have same contributions,
 we chose $P_{20}=1$ in this paper.
In the following subsection, we determine the remaining parity of $P_{15}$.

\subsection{Mass hierarchy between down type quark mass and charged lepton mass}

\begin{figure}[t]
 \centering
 \includegraphics[keepaspectratio, scale=0.7]{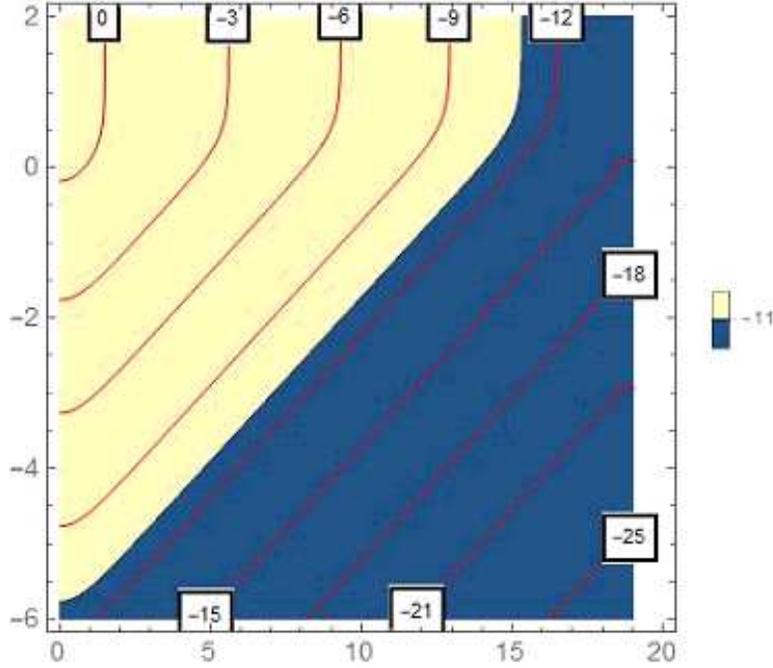}
 \caption{Common logarithms of the ratio of physical neutrino mass and W mass $m^{\nu}_\T{{phys}}/m_W$.
 The vertical axis is $\mbox{log}_{10}[\epsilon_6]$.
 The horizontal axis is $\lambda_{6}$.
 Parameters $c_1+c_2$, $\lambda_{15}$, $\epsilon_{15}$, $P_{15}$ and $P_{6}$ are taken to be $10$, $1$, $1$, $1$ and $1$, respectively.
 Red lines mean the order of magnitude for $m^{\nu}_\T{{phys}}/m_W$.
 Blue region is an allowed region for neutrino mass.}
 \label{neutrino_mass}
\end{figure}

In this subsection, we consider the mass hierarchy between down type quark mass and charged lepton mass.
According to section 3.1, the condition (\ref{VEV}) is needed.
Under this conditon, SM fermion mass (\ref{physmass}) and W boson mass (\ref{mw}) approximately proportional to VEV $\alpha$.
Therefore, the ratio of these masses does not depend on the Higgs VEV $\alpha$.

\begin{figure}[t]
 \centering
 \includegraphics[keepaspectratio, scale=1.0]{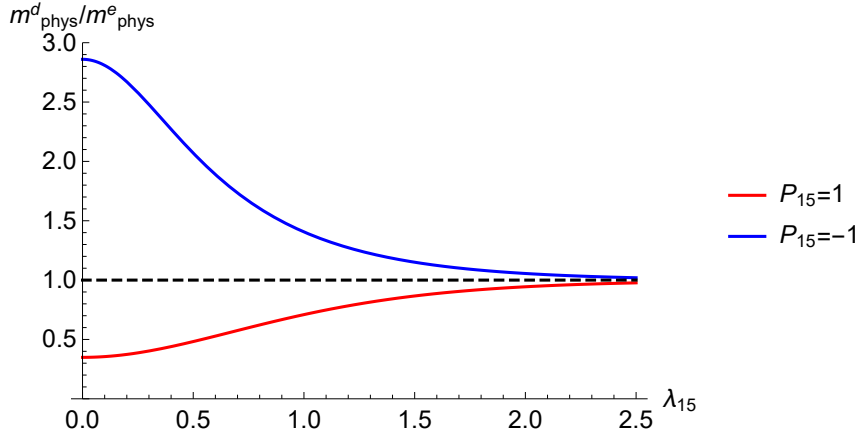}
 \caption{The ratio of the physical down-type mass and the physical charged lepton mass $m^d_\T{{phys}}/m^e_\T{{phys}}$.
 Parameters $\epsilon_{15}$, $\lambda_{20}$ and $\epsilon_{20}$ are taken to be $1$.}
 \label{hierarchy}
\end{figure}

The mass hierarchy depend heavily on the parameters
 $P_{15}$, $P_{6}$, $\lambda_{15}$, $\lambda_{6}$, $\epsilon_{15}$ and $\epsilon_{6}$.
Before the analysis of the mass hierarchy,
 we will obtain suitable regions for $\lambda_{6}$ and $\epsilon_{6}$
 by analyzing the physical neutrino mass $m^{\nu}_{\T{phys}}$.
The physical neutrino mass $m^{\nu}$ depend heavily on $\lambda_{6}$ and $\epsilon_{6}$.
Fig.~\ref{neutrino_mass} shows $\lambda_{6}$ and $\epsilon_{6}$ dependence on the neutrino mass $m^{\nu}$
 with $c_1+c_2=10$, $\lambda_{15}=1$, $\epsilon_{15}=1$, $P_{15}=1$ and $P_{6}=1$.
It is known that neutrino masses is smaller than 1 eV,
 so that the suitable region is $m^{\nu}_{\T{\mbox{phys}}}/m_W < 10^{-11}$
 (blue region in Fig.~\ref{neutrino_mass}).
In this region, contribution of the representation \B{6} are exponentially small,
 so that the contribution can be ignored to reproduce the SM fermion masses except for SM neutrinos.
In this case, the mass hierarchy is converge to 1
 by increasing $\lambda_{15}$ (Fig.~\ref{hierarchy}).
This figure indicates that
 down-type quark mass is smaller than charged leplon mass in the case of $P_{15}=1$,
 on the other hand,
 down-type quark mass is larger than charged leplon mass in the case of $P_{15}=-1$.
In SM, the hierarchies are
\begin{equation}
  \frac{m_{\T{\mbox{down}}}}{m_{\T{\mbox{electron}}}} \sim 9.1 >1,
  \frac{m_{\T{\mbox{strange}}}}{m_{\T{\mbox{muon}}}} \sim 0.9 <1,
  \frac{m_{\T{\mbox{bottom}}}}{m_{\T{\mbox{tauon}}}} \sim 2.3 >1,
\end{equation}
therefore the parity assignments of \B{15} representation for each generation have to be taken as
\begin{equation}
  P_{15}^{\T{\mbox{1st}}} = -1,
  P_{15}^{\T{\mbox{2nd}}} = 1,
  P_{15}^{\T{\mbox{3rd}}} = -1.
\end{equation}

\subsection{Reproducing top quark mass}
In our previous paper \cite{MY}, the up-type quark masses could not be larger than W boson mass,
although we had attempted some cases where top quark is embedded in higher rank representations
 whether the enhancement due to the group theoretical factor
 for the up-type quark masses can be obtained \cite{MY}.
As another possibility, it is known
 that the sizable localized gauge kinetic terms enhances fermion masses,
 which might be possible to reproduce top quark mass \cite{SSS}.
We consider this possibility in this paper.

\begin{figure}[t]
   \begin{tabular}{cc}
     \begin{minipage}[c]{0.47\hsize}
       \centering
       \includegraphics[keepaspectratio, scale=0.8]{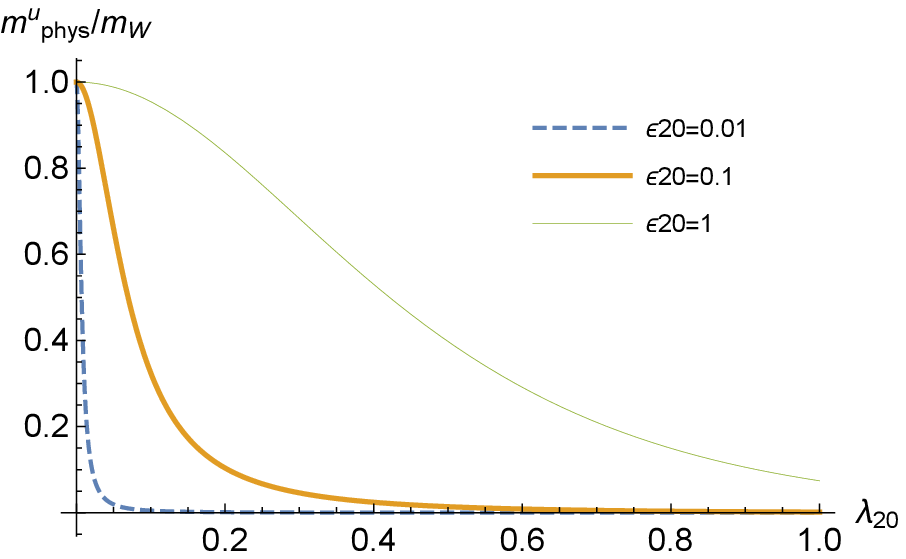}
     \end{minipage}
     &
     \begin{minipage}[c]{0.47\hsize}
       \centering
       \includegraphics[keepaspectratio, scale=0.8]{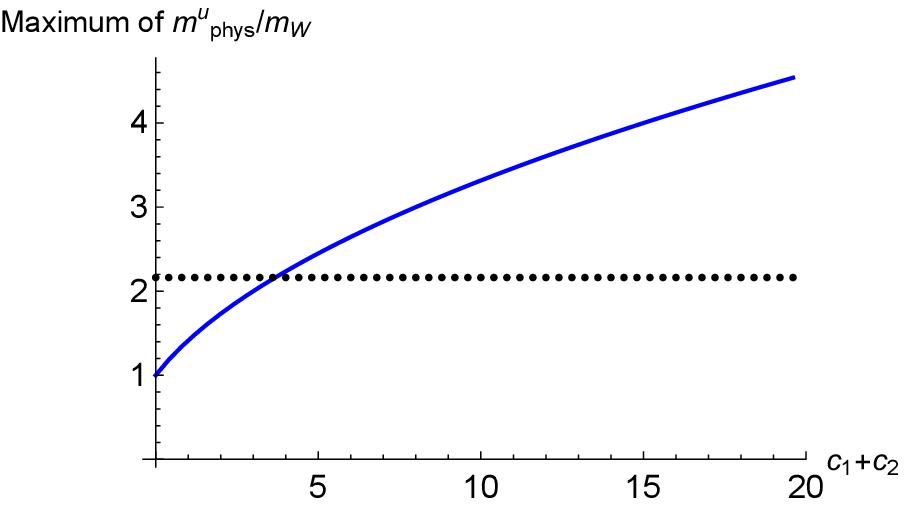}
     \end{minipage}\\
   \end{tabular}
  \caption{The bulk mass $\lambda_{20}$ dependence on the ratio of the physical up-type mass and W boson mass $m^u_\T{{phys}}/m_W$ with $c_1+c_2=0$ (left side) and the maximum ratio in the range of $0 \leq c_1 + c_2 \leq 20$ (right side).
  Other parameters are taken to be $1$.
  Dotted line means top quark mass: $m^u_\T{{phys}}/m_W=m_\T{{top}}/m_W\sim 2.15$.
  }
  \label{mlambda}
 \end{figure}

Fig.~\ref{mlambda} shows
 the bulk mass $\lambda_{20}$ dependence on
 the ratio of W boson mass and the physical up-type mass $m^u_\T{{phys}}$
 given by eq.~(\ref{physmass}).
To reproduce the top quark mass, 
 the maximum value of physical up-type quark mass
 has to be larger than the observed top quark mass 173 GeV.
We have studied the behavior of the maximum value in the range from $c_1 + c_2 = 0$ to $20$.
It turns out that
 the conditions where $c_1+c_2$ is at least larger than $4$
 is necessary to reproduce the top quark mass.

\begin{figure}[htb]
   \centering
   \includegraphics[keepaspectratio, scale=0.4]{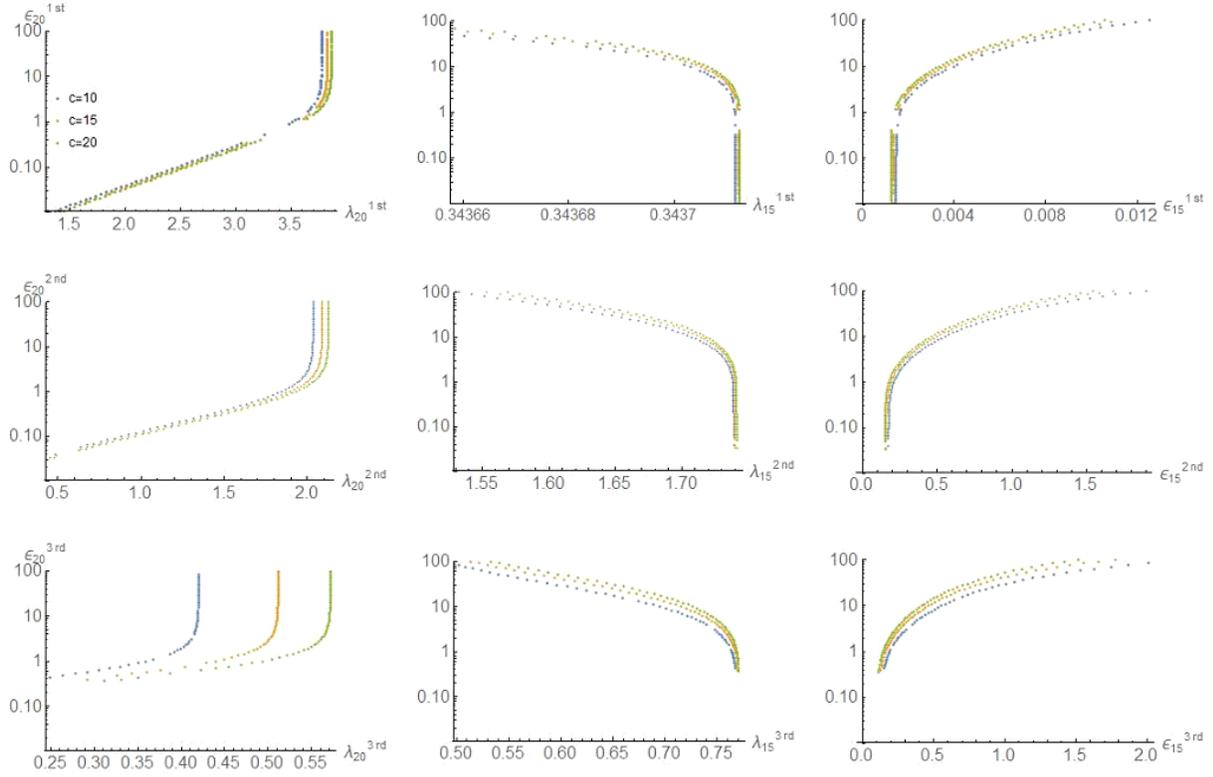}
   \caption{Scatter plots of the parameters reproducing the SM fermion masses except for neutrino masses.
   The upper, middle and bottom plots show $\epsilon_{20}$ dependences of $\lambda_{20}$, $\lambda_{15}$ and $\epsilon_{15}$ for the first, the second and the third generation, respectively.}
   \label{figure:fit_mass}
  \end{figure}

 Ignoring the contribution of representation \B{6}, three masses
  ($m^u_{\T{phys}}$, $m^d_{\T{phys}}$, $m^e_{\T{phys}}$)
  are fitted by 4 parametars
  ($\lambda_{20}$, $\epsilon_{20}$, $\lambda_{15}$, $\epsilon_{15}$),
  so that a degree of freedom is remaind to be free.
 Here, we try to get suitable parameters $\lambda_{20}$, $\lambda_{15}$, $\epsilon_{15}$
  by changing the value of $\epsilon_{20}$ at each generations (Fig.~\ref{figure:fit_mass}).
 We can find allowed parameter sets in a broad region.
 In the case of the first generation,
  the region that there is no point appears around $\epsilon_{20} \sim 0.8$
  since it is difficult to reproduce the mass hierarchy between the down quark and the electron,
  which is lager than those of the second and the third generation.
 In the case of the second and the third generations,
  the region that there is no point appears when $\epsilon_{20}$ is below a certain value
  since the up-type quark mass can not be reproduced.

 %

\section{Higgs effective potential}
In this section, we calculate the effective potential for the Higgs field
 and study whether the electroweak symmetry breaking correctly occurs.
Since the Higgs field is originally a gauge field,
 the potential is generated at one-loop by Coleman-Weinberg mechanism.
The potential from the bulk fields is given by
 \bea
 V(\alpha)
             =  \sum_n (\pm g) \int \frac{d^4p_E}{(2\pi)^4}
             \log [p_E^2 + m_n^2]
               \equiv g \mathcal{F}^\pm (q\alpha)
 \eea
 with
 \beq
 \mathcal{F}^{\pm}(q\alpha) = \pm \sum_n  \int \frac{d^4p_E}{(2\pi)^4}
 \log [p_E^2+m_n^2],
 \eeq
where overall signs $+(-)$ stand for fermion (boson) contributions, respectively.
$g$ means the spin degrees of freedom of the field running in the loop.
The loop momentum $p_E$ is taken to be Euclidean.

For bulk fermions and mirror fermions,
 the mass spectrum is calculated as the following four types of form
 depending on the $Z_2$ parity and the bulk mass.
 \bea
 && m_n^2 = \frac{(n + q \alpha)^2}{R^2}, \non
 && m_n^2 = \frac{(n + 1/2 + q \alpha)^2}{R^2}, \non
 && m_n^2 = \frac{(n + q \alpha)^2}{R^2} + \left( \frac{\lambda}{\pi R} \right)^2, \non
 && m_n^2 = \frac{(n + 1/2 + q \alpha)^2}{R^2} + \left( \frac{\lambda}{\pi R} \right)^2.
 \eea
The first (second) half of spectrum correspond to the spectrum of massless (massive) bulk fields.
The first and third (the second and the last) types of spectrum correspond to
 the spectrum of the fields with (anti-)periodic boundary conditions.
Using this information, we obtain the corresponding potentials \cite{CCP}.
 \bea
 \mathcal{F}^{\pm}(q\alpha) &=&
        \mp \frac{3}{64\pi^6 R^4} \sum_{k=1}^{\infty} \frac{\cos (2\pi q\alpha k)}{k^5}, \non
 \mathcal{F}^{\pm}_{1/2}(q\alpha) &=&
        \mp \frac{3}{64\pi^6 R^4} \sum_{k=1}^{\infty}(-1)^k \frac{\cos (2\pi q \alpha k)}{k^5}, \non
 \mathcal{F}^{\pm}_\lambda (q\alpha)&=&
       \mp \frac{3}{64\pi^6 R^4} \sum_{k=1}^{\infty}
       \frac{\cos(2\pi q \alpha k) e^{-2 k \lambda}} {k^3}
       \left[\frac{(2\lambda)^3}{3} + \frac{2\lambda}{k}+\frac{1}{k^2}\right],\non
 \mathcal{F}^{\pm}_{1/2\lambda}(q\alpha)&=&
       \mp \frac{3}{64\pi^6 R^4}\sum_{k=1}^{\infty}
       (-1)^k\frac{\cos(2\pi q \alpha k)e^{-2k \lambda}} {k^3}
       \left[\frac{(2\lambda)^3}{3}+\frac{2\lambda}{k}+\frac{1}{k^2}\right].
 \eea

\noindent
Table~\ref{table:potential} lists the various potentials
 from bulk fermion and mirror fermion contributions.
The coefficients in front of the each potential can be read
 from the branching rules in the decomposition of the $SU(6)$ representation
 into $SU(3)_C \times SU(2)_L \times U(1)_Y \times U(1)_X$ representations
 listed in Appendix A of our previous paper \cite{MY}.

 \begin{table}[t]
  \centering
   \begin{tabular}{|c||c|} \hline
   bulk+mirror
      &$g=8$      \\ \hline
   $20^{(+,+)}+20^{(-,-)}$
      & $3\mathcal{F}^-_{\lambda}(\alpha)
         +3\mathcal{F}^-_{1/2\lambda}(\alpha)$ \\ \hline
   $15^{(+,+)} +15^{(-,-)}$
      &$\mathcal{F}^-_{\lambda}(\alpha)
       +3\mathcal{F}^-_{1/2\lambda}(\alpha)$        \\ \hline
   $15^{(+,-)} +15^{(-,+)}$
      &$3\mathcal{F}^-_{\lambda}(\alpha)
       +\mathcal{F}^-_{1/2\lambda}(\alpha)$        \\ \hline
   $6^{(-,+)} +6^{(+,-)}$
      &$\mathcal{F}^-_{1/2\lambda}(\alpha)$       \\ \hline
   $6^{(-,-)} +6^{(+,+)}$
      &$\mathcal{F}^-_{\lambda}(\alpha)$       \\ \hline
  \end{tabular}
  \caption{Bulk fermion and mirror fermion contribution to Higgs potential.}
  \label{table:potential}
 \end{table}

Next, we calculate the gauge field loop contributions to the effective potential.
As was mentioned in Section 3,
 the gauge boson mass spectrum is complicated because of the localized gauge kinetic terms.
Here, we derive the effective potential without solving explicit KK mass spectrum of gauge fields.
For the quantization of the gauge fields,
 we fix the gauge by using the background field method
 where the gauge fields are divided into the classical field $\overline{\C{\C{A}}}_M$
 and quantum field $\C{\C{A}}_M$.
The gauge fixing condition function is given
 by $G^a(x) =\overline{D}_M^{ab}\C{\C{A}}^{M\,b}(x) - \omega^a(x)$ as usual,
 where $\overline{\C{\C{A}}}_M = \delta_{My}\frac{2\alpha}{Rg}T^{28}$,
  $\overline{D}^{M\,ab}$ is covariant derivative containing a classical field only
  $\overline{D}^{M\,ab} \C{\C{A}}^b_M = (\delta^{ab}\del^M -g_5 f_0^{abc}\overline{\C{\C{A}}}^{M\,c})\C{\C{A}}_M^b$
  and $\omega(x)$ is an arbitrary scalar function.

We note that the contributions from the gauge fields are different depending on the boundary conditions.
In the periodic sector,
 the quadratic terms in eq.~(\ref{lag_g}) become
 \bea
 \C{L}_{\T{ quadratic}} &=&
 \frac{1}{2}\mathcal{\C{A}}^{a,\mu}g_{\mu\nu}(\overline{D}_{P}\overline{D}^{P})^{ab}\mathcal{\C{A}}^{b,\nu}
 +\frac{1}{2}\mathcal{\C{A}}^a_y(\overline{D}_{P}\overline{D}^{P})^{ab} \mathcal{\C{A}}^b_y
 + \overline{\B{c}}^a\overline{D}^{M,ab}\overline{D}^{bc}_{M}\B{c}^c  \nonumber \\
 &&
 +[2\pi R c_1\delta(y)+2\pi R c_2\delta(y-\pi R)]\frac{1}{2}\mathcal{\C{A}}^{a,\mu} \delta^{ab}
 (g_{\mu\nu}\overline{D}_{\rho}\overline{D}^{\rho}-\overline{D}_{\mu}\overline{D}_{\nu})\mathcal{\C{A}}^{b,\nu},
 \eea
 where $\B{c}$ ($\overline{\B{c}}$) denotes the ghost (anti-ghost) field.
After the KK expansion of the gauge and the ghost fields
 and diagonalizing 4D gauge kinetic terms, the contribution to Higgs potential can be written down as
 \beq
 \int \frac{d^4 p}{(2\pi)^4} \frac{1}{2}\log \det\left[\frac{K^{\C{A}_M}}{(K^{\T{ ghost}})^2}\right],
 \label{potential from gauge}
 \eeq
 where
 \bea
 K^{\C{A}_5,\T{ ghost}}_{mn} &=& \delta_{mn}(p^2+m_n^2), \\
    K^{\C{A}_{\mu}}_{mn,\mu\nu} &=& \delta_{mn}g_{\mu\nu}(p^2+m_n^2)
        +(c_1+(-1)^{m+n}c_2)(g_{\mu\nu}p^2-p_{\mu}p_{\nu}).
 \eea
Using the following determinant results
 \bea
 &&\det_{(\mu\nu)}\left[\delta_{nm}g_{\mu\nu}(p^2+m_n^2)+(c_1+(-1)^{m+n}c_2)(g_{\mu\nu}p^2-p_{\mu}p_{\nu})\right] \nonumber \\
 &&\hspace{30pt}= -\delta_{nm}(p^2+m_n^2)\left[\delta_{nm}(p^2+m_n^2)+(c_1+(-1)^{m+n}c_2)p^2\right]^3, \\
  &&\det_{(nm)}\left[\delta_{nm}(p^2+m_n^2)+(c_1+(-1)^{m+n}c_2)p^2\right] \nonumber \\
  &&\hspace{30pt}= \Pi_n(p^2+m_n^2)\left[ \Pi_{i=1}^2 \left(1+c_i \sum_n \frac{p^2}{p^2+m_n^2}\right)
                 -\Pi_{i=1}^2 \left(c_i \sum_n \frac{p^2(-1)^n}{p^2+m_n^2}\right)\right]
 \eea
 where $(\mu\nu)$ denots the determinant over 4D spacetime
 and $(nm)$ denotes the determinant over the KK mode,
 eq.~(\ref{potential from gauge}) is computed as follows.
\beq
\int \frac{d^4 p}{(2\pi)^4} \frac{1}{2}\log \det\left[\frac{K^{\C{A}_M}}{(K^{gh})^2}\right]
=\mathcal{F}^+(q\alpha)+\mathcal{F}^c(q\alpha),
\eeq
\beq
\mathcal{F}^c(q\alpha)
     = \frac{3}{16\pi^6 R^4}\int dx \, x^3 \log \left[\Pi_{i=1}^2 \left(1+c_i \sum_n x \mbox{Re}f_0^{(+)}(x,q\alpha) \right)
              -\Pi_{i=1}^2 \left(
              c_i \sum_n x \mbox{Re}f_1(x,q\alpha) \right)
              \right]
\label{periodic  potential}
\eeq
with
\beq
f_1(x,q\alpha)
= \sum_{n=-\infty}^{\infty}\frac{(-1)^n}{|x|+i\pi(n+q\alpha)}
=\sinh^{-1}(|x|+i\pi q\alpha).
\eeq
$\mathcal{F}^+(q\alpha)$ and $\mathcal{F}^c(q\alpha)$ are contributions
 from the bulk gauge kinetic terms and the localized gauge kinetic terms, respectively.
It is easy to check $\mathcal{F}^c(q\alpha)=0$ at $c_i=0$.

In the anti-periodic sector,
 a difference from the periodic sector is the following.
 \beq
 K^{\C{A}_{\mu}}_{mn,\mu\nu} = \delta_{mn}g_{\mu\nu}(p^2+m_n^2)
     +c_1(g_{\mu\nu}p^2-p_{\mu}p_{\nu}).
 \eeq
Therefore, substituting $c_2=0$ in eq.~(\ref{periodic  potential}),
 we easily obtain contributions from the localized gauge kinetic terms with anti-periodic boundary condition.
\beq
\mathcal{F}_{1/2}^c(q\alpha)
     = \frac{3}{16\pi^6 R^4}\int dx \, x^3 \log \left[1+c_1 \sum_n x \mbox{Re}f_0^{(-)}(x,q\alpha)\right].
\eeq
Table~\ref{table:potential gauge} lists a Higgs potential
 from gauge field contributions.
 \begin{table}[t]
  \centering
   \begin{tabular}{|c||c|} \hline
   gauge
      &$g=3$      \\ \hline
   $35^{(+,+)}$
      & $2\mathcal{F}^+(\alpha)
         +\mathcal{F}^+(2\alpha)
         +6\mathcal{F}^+_{1/2}(\alpha)
         +2\mathcal{F}^c(\alpha)
         +\mathcal{F}^c(\alpha)
         +6\mathcal{F}^c_{1/2}(\alpha)$ \\ \hline
  \end{tabular}
  \caption{Gauge field contributions to Higgs potential.}
  \label{table:potential gauge}
 \end{table}


Finally, we need the contributions from the SM fermion localized at $y=0$ to the Higgs potential.
The results are obtained from the expression below \cite{MY}.
\begin{equation}
  V_{a} = -\frac{1}{4\pi^6 R^4}
          \int dx \, x^3
          \log \left[Z_L^a Z_R^a +\frac{\pi R}{x}m^a\right] \,
  (a=u,d,e,\nu).
\end{equation}

In calculation of the potential from the both bulk and boundary contributions,
 we have subtracted the $\alpha$ independent part of the potential
 since it corresponds to the divergent vacuum energy
 and is irrelevant to the electroweak symmetry breaking.

\begin{figure}[t]
 \begin{tabular}{cc}
 \begin{minipage}[c]{0.44\hsize}
 \centering
 \includegraphics[keepaspectratio, scale=0.7]{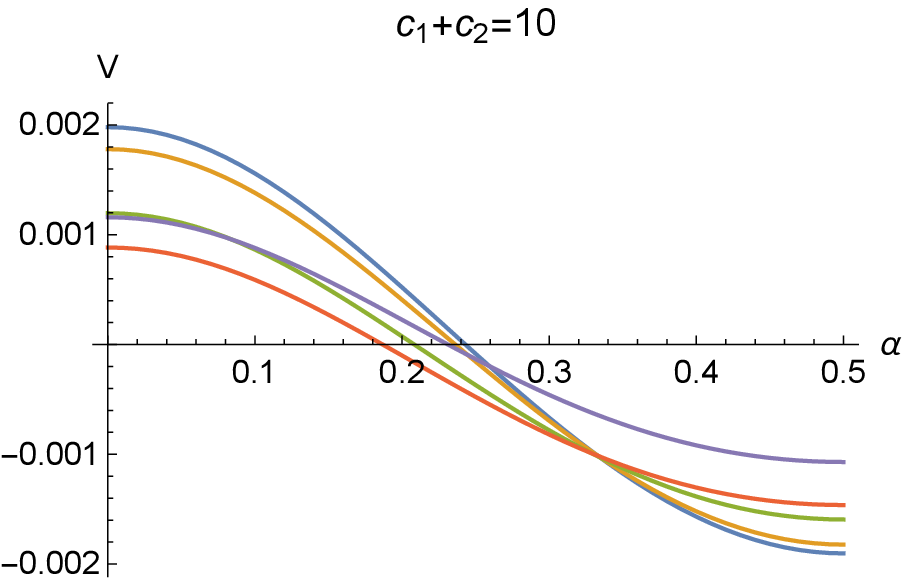}
 \end{minipage}&
 \begin{minipage}[c]{0.5\hsize}
 \centering
 \includegraphics[keepaspectratio, scale=0.7]{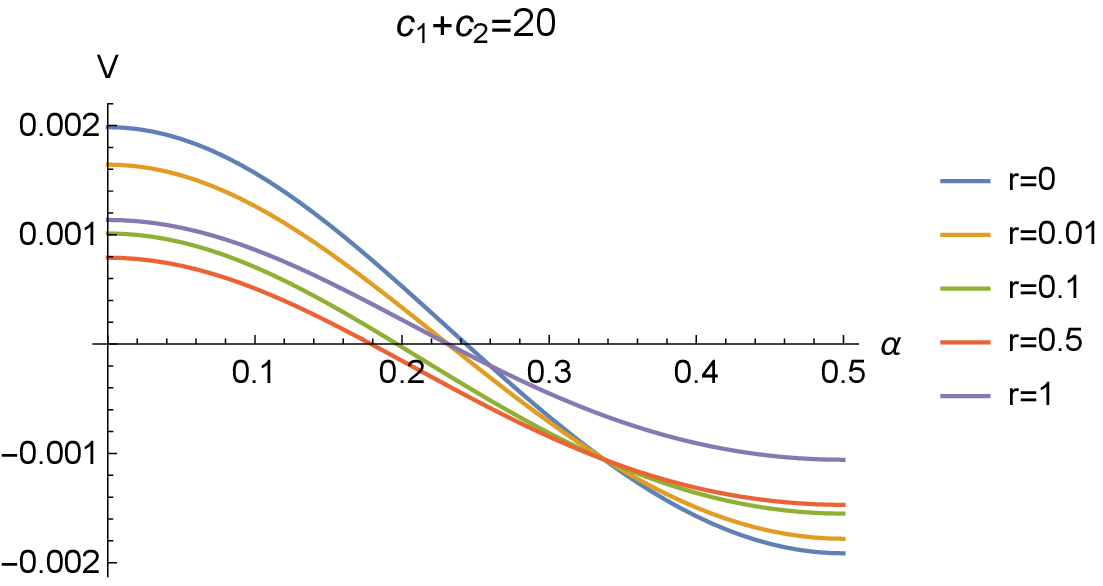}
 \end{minipage}\\
 \end{tabular}
 \caption{Total Higgs potential in the case of $c_1+c_2=10$ (left side) and
 $20$ (right side).
 The horizontal axis is the VEV of Higgs field.}
 \label{figure:noextra}
\end{figure}

Total potential is $V(\alpha)=V_{\T{ gauge}}+V_{\T{ bulk}}+V_{\T{ boundary}}$,
 where $V_{\T{ gauge}}$, $V_{\T{ bulk}}$ and $V_{\T{ boundary}}$ are
 the contributions from the gauge field, the bulk and mirror fermions
 and the mixing between the bulk fermions and the SM fermions respectively.
In this model, \B{6} bulk mass $\lambda_6$ is lager than other bulk muss
 to reprodce small neutrino mass as mentioned in section 4.2,
 so that \B{6} bulk and mirror fermion contribution to Higgs potential
 and mixing contribution $V_{\nu}$ are negligibly small.
As a result,
 our analysis is independent of parameters $\lambda_6$, $\epsilon_6$ and $P_{6}$.
The plots of the potential are shown in Fig.~\ref{figure:noextra}.
It is reasonable to fix a sum $c_1+c_2$ to constant value $c = c_1 + c_2$
 since the bulk mass $\lambda$ depends on $c$. 
Then, the behavior of potential can be considered
 by changing a ratio $r=c_1/c=1-c_2/c$ $(0\leq r \leq 1)$.
From the requirement of large compactification scale in Section 3.1,
 $\alpha$ has to be smaller than 0.1.
This implies that the electroweak symmetry is not broken in our model as it stands
 since the minimum is $\alpha=0.5$.
Therefore, we need to extend our model
 and introduce extra fermions to obtain $0 < \alpha < 0.1$
 for a successful electroweak symmetry breaking.

\begin{figure}[t]
 \begin{tabular}{cc}
 \begin{minipage}[c]{0.47\hsize}
 \centering
 \includegraphics[keepaspectratio, scale=0.8]{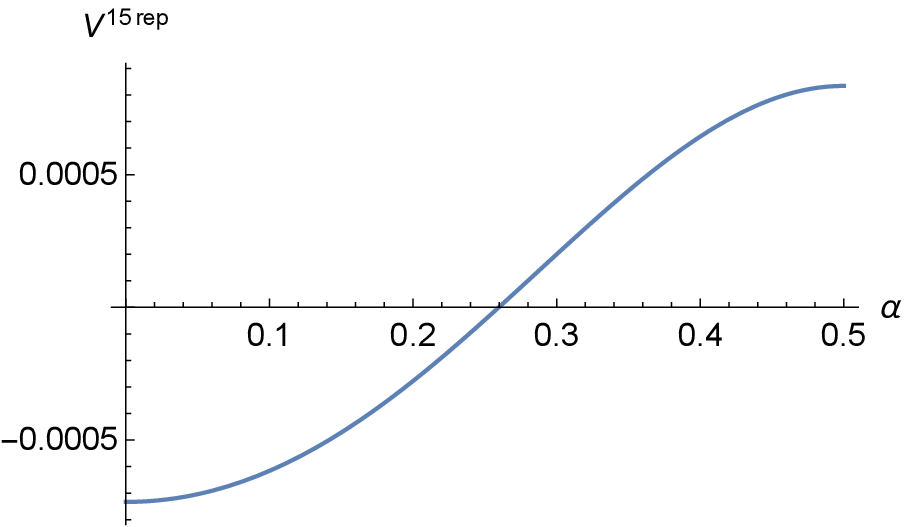}
 \end{minipage}&
 \begin{minipage}[c]{0.47\hsize}
 \centering
 \includegraphics[keepaspectratio, scale=0.8]{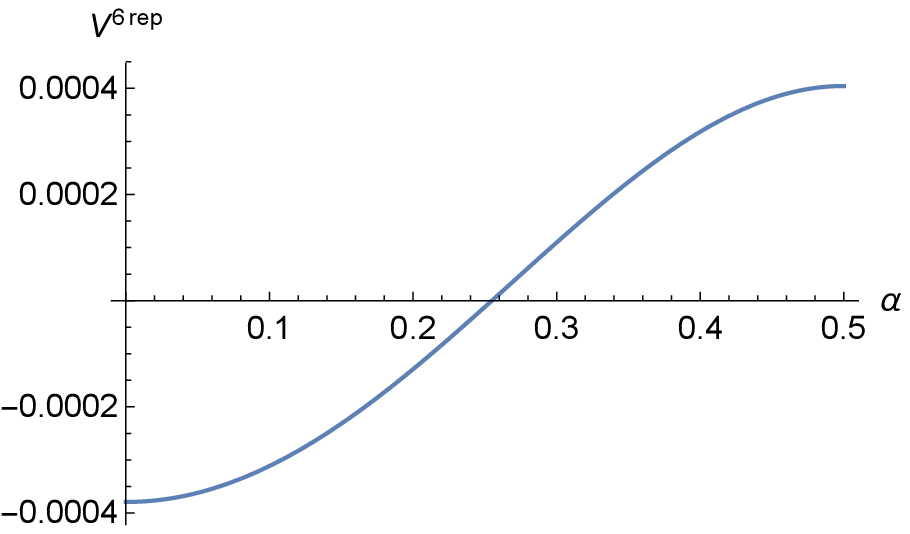}
 \end{minipage}\\
 \end{tabular}
 \caption{Potential for a set of \B{15} periodic bulk and mirror fermions with extra bulk mass $\lambda_\T{{ext}}=0$(left side) and potential for a set of \B{6} periodic bulk and mirror fermions with extra bulk mass $\lambda_\T{{ext}}=0$(right side).}
 \label{figure:156tpotential}
\end{figure}

In this paper,
 we introduce
 three sets of \B{15} or five sets of \B{6} periodic bulk and mirror fermions.
In Table~\ref{table:potential},
 the contribution of a set of \B{15} and \B{6} periodic bulk and mirror fermions to Higgs potential
 are given
 and its plot of the potential is shown in Fig.~\ref{figure:156tpotential}.
The strategy of introducing such a set of bulk and mirror fermions is as follows.
Since the total potential without extra fermions has a minimum at $\alpha=0.5$,
 the contribution of extra potential with a minimum at $\alpha < 0.1$ is needed.
As can be seen from Fig.~\ref{figure:156tpotential},
 the contribution from massless \B{15} fermions and \B{6} fermions to the potential has a minimum at $\alpha=0$
 where the correct electroweak symmetry breaking does not occur.
On the other hand, the contribution of massive \B{15} and \B{6} is suppressed by the bulk mass $\lambda_\T{{ext}}$,
 then $\alpha$ in the total potential with extra matter contributions
 becomes small by increasing $\lambda_\T{{ext}}$.
This opens a possibility that $0 < \alpha < 0.1$ required
 for the correct pattern of the electroweak symmetry breaking can be realized.

\begin{figure}[t]
  \begin{tabular}{c}
  \begin{minipage}[c]{0.9\hsize}
  \centering
  \includegraphics[keepaspectratio, scale=1]{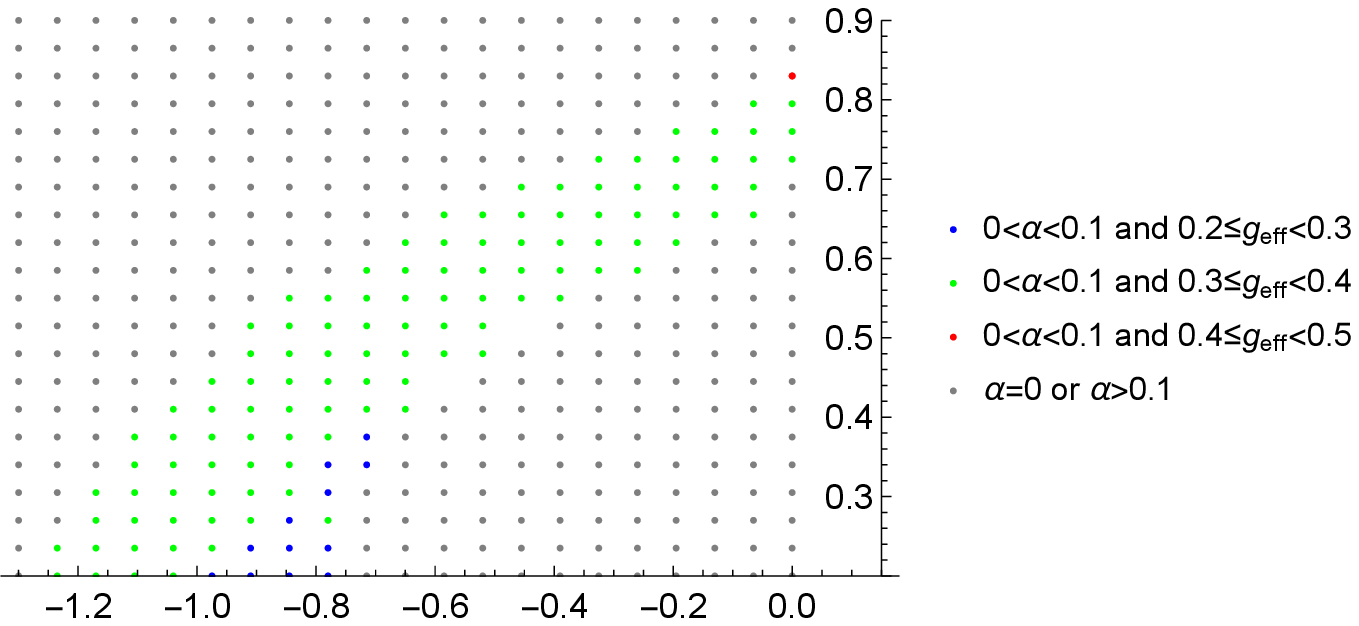}
  \end{minipage}\\
  \\
  \begin{minipage}[c]{0.9\hsize}
  \centering
  \includegraphics[keepaspectratio, scale=1]{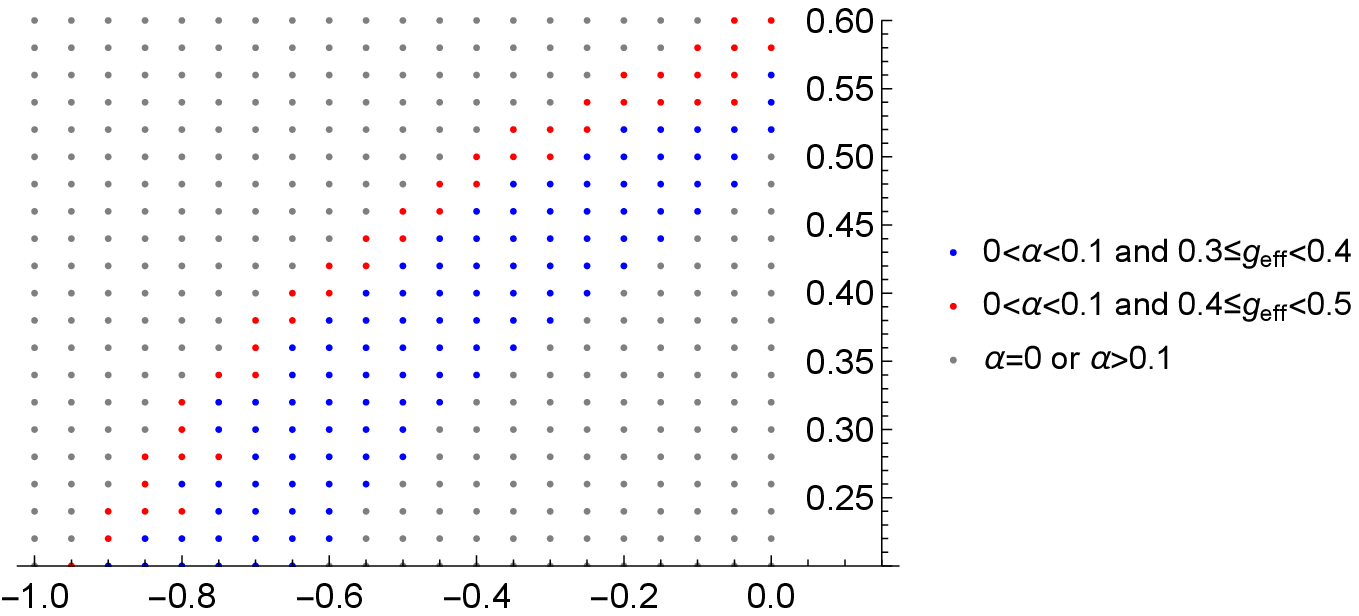}
  \end{minipage}\\
  \end{tabular}
  \caption{Bulk mass of extra fermion and $r$ dependence for the 4D gauge coupling
  for introduced three sets of \B{15} bulk fermion (upper side)
  and five sets of \B{6} bulk fermion (lower side).
  Horizontal and vertical axis are $\log_{10}[r]$ and $\lambda_{\T{ext}}$ respectively.}
  \label{figure:withextra}
 \end{figure}

Higgs mass is obtained from the second derivative of total potential as
 \beq
 m_H = \frac{g_{{\rm eff}} R \sqrt{1+c}}{2}\sqrt{V''(\alpha)} \sim \mbox{125 GeV}.
 \eeq
The compactification scale $1/R$ is defined by eq.~(\ref{mw})
 \beq
 \frac{1}{R}=\frac{\pi\sqrt{1+c}\times 80.3\mbox{ GeV}}{\sin(\pi \alpha)}.
 \eeq
The compactification scale can be large
 by increasing $c$ and decreasing  $\alpha$ (or equivalently increasing $\lambda_\T{{ext}}$).
Fig.~\ref{figure:withextra} shows a plot of 4D gauge coupling $g_\T{ eff}$
 for $-1.3 < \log_{10}[r] < 1$ and $0.20 \leq \lambda_\T{{ext}} \leq 0.90$
 in the case of three sets of \B{15} fermion
 and
 for $-1.0 < \log_{10}[r] < 0$ and $0.20 \leq \lambda_\T{{ext}} \leq 0.60$
 in the case of five sets of \B{6} fermion.
In this plot, $\alpha$ and the compactification scale are also displayed.
We can find an allowed region of parameter space in our model,
 where the gauge coupling universality is kept ($\alpha < 0.1$),
 the top quark ($c > 4$), Higgs boson masses and the realistic electroweak symmetry breaking ($0<\alpha<0.5$) are obtained.
As representative samples of our solutions, we list
 $g_\T{ eff}$ and $1/R$ are 0.349 and 8 TeV
 with $\lambda_\T{{ext}}=0.71$, $r=1$, $c=7$ and $\epsilon_{20}^i$=(0.1,1,1)
 in the case of three sets of \B{15} fermion
 and
 $g_\T{ eff}$ and $1/R$ are 0.371 and 16.2 TeV
 with $\lambda_\T{{ext}}=0.71$, $r=1$, $c=7$ and $\epsilon_{20}^i$=(0.1,1,1)
 in the case of three sets of \B{6} fermion
Note that the compactification scale in the present paper becomes larger
 by the effects of the localized gauge kinetic terms,
 which is compared to the slightly small compactification scale ($\sim 0.8$ TeV) in our previous paper \cite{MY}.

We give some comments on the extra bulk fermions
 which are required for the realistic electroweak symmetry breaking.
First, their representations are very simplified.
Although the representation is a totally four rank symmetric tensor of $SU(6)$ in our previous paper,
 the representation is at most two rank symmetric tensor in the present analysis.
Second, it is natural to ask whether there are any allowed region of parameters
 in cases with one or two sets of extra fermions in the \B{15} representation.
We have also tried the potential analysis for those cases,
 but could not obtain an observed Higgs boson mass.
The same is true for \B{6} representation.

\section{Conclusions}
In this paper, we have discussed the fermion mass hierarchy
 in $SU(6)$ GGHU with localized gauge kinetic terms.
The SM fermions are introduced in the $SU(5)$ multiplets on the boundary at $y=0$.
We also introduced massive bulk fermions in three types of $SU(6)$ representations
 coupling to the SM fermions on the boundary.
Once the localized gauge kinetic terms are present,
 the zero mode wave functions are distorted and the gauge coupling universality is not guaranteed.
We have investigated the constraints where the gauge coupling constant between the SM fermions and a SM gauge field,
 the cubic and the quartic self-interaction gauge coupling constants are almost universal.
It turns out that the gauge coupling universality can be preserved
 if the dimensionless Higgs VEV is smaller than 0.1 ($\alpha < 0.1$).

We have shown that the SM fermion masses including top quark can be reproduced
 by mild tuning of bulk masses and the parameters of the localized gauge kinetic terms.
As for top quark mass, we have investigated a dependence of the maximum of fermion mass
 on a parameter $c$ of the localized gauge kinetic terms.
 $c > 4$ is found for obtaining top quark mass in Fig.~\ref{mlambda}

We have also calculated additional contributions
 to one-loop Higgs potentials from the localized gauge kinetic terms.
It was found as in our previous paper \cite{MY}
 that the electroweak symmetry breaking does not occur
 for the fermion matter content mentioned above.
To overcome this problem,
 we have shown that the electroweak symmetry breaking happened
 by introducing additional three sets of bulk and mirror fermions in ${\bf 15}$ representation
 or five sets of bulk and mirror fermions in ${\bf 6}$ representation.
Note that the representation was very simplified comparing to our previous case
 where it was the ${\bf 126}$ representation.
The effects of localized gauge kinetic terms enhanced the compactification scale,
 which is compared to the small compactification scale
 ($\sim 16$ TeV) in our previous paper \cite{MY}.
This enhancement of the compactification scale also helps Higgs boson mass large.
The observed SM Higgs boson mass 125 GeV was indeed obtained in our analysis.

There are issues to be explored in a context of GUT scenario.
First one is the gauge coupling unification.
It is well known that the gauge coupling running in (flat) extra dimensions
 is the power dependence on energy scale \cite{DDG} not the logarithmic one.
Therefore, the GUT scale is likely to be very small comparing to the conventional 4D GUT.
It is therefore very nontrivial whether the unified $SU(6)$ gauge coupling
 at the GUT scale is perturbative
 since many bulk fields are introduced,
 which might lead to Landau pole below the GUT scale.
Second one is proton decay.
$X, Y$ gauge boson masses are likely to be extremely light comparing to the conventional GUT scale.
Therefore, proton decays very rapidly and our model is immediately excluded
 by the constraints from the Super Kamiokande data as it stands.
Dangerous baryon number violating operators must be forbidden
 by some symmetry (see \cite{PDUED} for UED case) for the proton stability.
If $U(1)_X$ is broken to some discrete symmetry which plays an role for it,
 it would be very interesting.

These issues are remained for our future work.

\section*{Acknowledgments}
This work is supported in part by JSPS KAKENHI Grant Number JP17K05420 (N.M.).

\appendix




  \end{document}